# Rates of Stellar Tidal Disruption Events around Intermediate-mass Black Holes

Janet N. Y. Chang[1], Lixin Dai[1], Hugo Pfister[1], Rudrani Kar Chowdhury[1], and Priyamvada Natarajan[2,3,4]
[1] Department of Physics, The University of Hong Kong, Pokfulam Road, Hong Kong; lixindai@hku.hk
[2] Department of Astronomy, Yale University, New Haven, CT 06511, USA
[3] Department of Physics, Yale University, New Haven, CT 06520, USA
[4] Black Hole Initiative at Harvard University, 20 Garden St., Cambridge, MA 02138, USA

## Abstract

Rates of stellar tidal disruption events (TDEs) around supermassive black holes (SMBHs) have been extensively calculated using the loss cone theory, while theoretical work on TDE rates around intermediate-mass black holes (IMBHs) has been lacking. In this work, we aim to accurately calculate the IMBH TDE rates based on their black hole (BH) masses and the stellar profiles of their host galaxies obtained from the latest observations. We find that the TDE rate per galaxy for IMBHs in the center of small galaxies is similar to that of SMBH TDEs, while the TDE rate per cluster from IMBHs in globular clusters is much lower. Very interestingly, we show that the rate of IMBH TDEs generally increases with the BH mass, which is opposite to the trend seen in SMBH TDEs. As a result, the volumetric TDE rate peaks around a BH mass of $10^6\,M_\odot$. The IMBH TDEs from galactic nuclei have an overall volumetric rate comparable to SMBH TDEs at $\sim 10^{-7}\,\mathrm{Mpc}^{-3}\,\mathrm{yr}^{-1}$, and off-center IMBH TDEs from globular clusters have a volumetric rate that is one or two orders of magnitude lower, assuming that their occupation fraction varies within 10%–100%. Furthermore, we report that IMBH TDEs typically occur in the pinhole regime, which means that deeply plunging events are more likely for IMBH TDEs compared to SMBH TDEs.

*Unified Astronomy Thesaurus concepts:* Tidal disruption (1696); Intermediate-mass black holes (816); Galactic center (565); Globular star clusters (656); Stellar dynamics (1596)

## 1. Introduction

The study of intermediate-mass black holes (IMBHs), with masses that lie between those of stellar-mass black holes (BHs; $<100\,M_\odot$) and supermassive BHs (SMBHs; $>10^6\,M_\odot$), has sparked a lot of recent interest (see review by J. E. Greene et al. 2020). One important reason for studying IMBHs is that they provide insights into the formation and evolution of SMBHs (J. R. Bond et al. 1984; S. F. Portegies Zwart & S. L. W. McMillan 2002; V. Bromm & A. Loeb 2003; S. F. Portegies Zwart et al. 2004; G. Lodato & P. Natarajan 2006), leading to significantly different predictions for the IMBH number density, occupation fraction (the fraction of galaxies hosting a massive BH (MBH) in the center), and BH−galaxy scaling relations. Therefore, understanding the IMBH population is important, as they serve as the bridge to the SMBH population. Additionally, in the Romulus and ASTRID cosmological simulations, in which central BHs are not forcefully pinned to the centers of their parent halos and host galaxies, a significant population of wandering IMBHs are predicted in most, if not all, galaxies (A. Ricarte et al. 2021b; T. Di Matteo et al. 2023). Therefore, observationally, we currently do not have a full census of the IMBH population.

Direct detection of IMBHs and measurements of their masses are rather challenging owing to the limitations of current techniques. One of the most accurate methods is dynamical measurements, which are done via measuring the motions of stars or gases within the BH's sphere of influence. However, this technique requires very high resolution observations and therefore can only be applied to nearby galaxies, which are up to a few Mpc away (J. Gerssen et al. 2002; A. J. Barth et al. 2009; E. Noyola et al. 2010; N. Neumayer & C. J. Walcher 2012; N. Lützgendorf et al. 2013, 2015; M. den Brok et al. 2015; D. D. Nguyen et al. 2017, 2019; B. Kızıltan et al. 2017; B. B. P. Perera et al. 2017; H. Baumgardt 2017). Another method for detecting and estimating IMBH masses is through their accretion luminosity. However, since accretion luminosity generally scales with the BH mass $M_{\rm BH}$, with the low luminosities of IMBHs it is difficult to discriminate against the emission produced from young starbursts (J. M. Cann et al. 2019). As a result of these observational constraints, only a limited number (around 20–40) of IMBH candidates have been detected so far (J. E. Greene et al. 2020).

Recently, it has been proposed that tidal disruption events (TDEs) can provide a unique probe of IMBHs. TDEs are produced when a star ventures too close to a BH and gets ripped apart by extreme tidal forces (M. J. Rees 1988). Following the disruption, a large part of the stellar debris remains bound to the BH, and a luminous flare is produced through the collision or accretion of these debris. TDEs, it has been demonstrated, can help us constrain BH mass and spin (M. Kesden 2012; G. Leloudas et al. 2016; B. Mockler et al. 2019) and study BH accretion and outflow physics (J. L. Dai et al. 2021). Furthermore, since the disrupted stars are scattered into the "loss cone" through two-body interactions with other stars, the TDE rate can be used to probe the stellar population, dynamics, and structure in the inner part of the host galaxy (J. Magorrian & S. Tremaine 1999; J. Wang & D. Merritt 2004).

Extensive theoretical calculations of TDE rates have been carried out. J. Magorrian & S. Tremaine (1999) and J. Wang & D. Merritt (2004) showed that, assuming a simple isothermal stellar density profile and constant velocity dispersion, the typical TDE rates for SMBHs are around $10^{-4}$ to $10^{-3}\,\mathrm{galaxy}^{-1}\,\mathrm{yr}^{-1}$, with a weak dependence on the BH mass. Recently, more accurate studies of the TDE rates around SMBHs were conducted by N. C. Stone & B. D. Metzger (2016) and H. Pfister et al. (2020), who employed more realistic stellar profiles from observations of nearby, massive galaxies. The TDE rates from these galaxies can range between $10^{-8}$ and $10^{-3}$ galaxy$^{-1}$ yr$^{-1}$, and the large variation is introduced owing to the differences in the host stellar structure. Furthermore, H. Pfister et al. (2020) found that the TDE rates in galaxies with nuclear star clusters (NSCs) can be enhanced by up to 2 orders of magnitude. These NSCs are clusters of stars that reside in the center of galaxies, which are commonly found in galaxies with stellar masses less than $10^{10} M_\odot$ (N. Neumayer et al. 2020). All of the works referenced above have focused on calculating the TDE rates in the SMBH regime, and they all find that the expected TDE rate from a galaxy on average declines with increasing mass of the SMBH or host galaxy.

Observationally, about 100 TDEs have been detected so far (S. Gezari 2021), a few of which are likely produced from IMBHs (W. P. Maksym et al. 2013; T. Wevers et al. 2017; D. Lin et al. 2018; C. R. Angus et al. 2022). IMBH TDEs are expected to generate different observational signatures compared to SMBH TDEs. Since the debris fallback time scales with $M_{\rm BH}^{1/2}$ (J. Guillochon & E. Ramirez-Ruiz 2013), IMBH TDE flares have been suggested to have faster evolution timescales. For a solar-type star TDE, the fallback timescale ranges from weeks to months for an SMBH but shortens to be only a few days for an IMBH. Therefore, it has been proposed that fast blue optical transients with rapid rise times could be powered by IMBH TDEs (R. Margutti et al. 2019; N. P. M. Kuin et al. 2019; D. A. Perley et al. 2019). On the other hand, stellar debris can circularize very slowly around IMBHs since the stream self-collisions are weak (L. Dai et al. 2015; H. Shiokawa et al. 2015; T. H. T. Wong et al. 2022), which can possibly greatly lengthen the TDE flare rise timescale. More theoretical modeling is needed to check which effect dominates and whether TDE flares do evolve more quickly around IMBHs than SMBHs. In addition, IMBH TDEs should be dimmer than SMBH TDEs if their accretion power is Eddington limited, although recent numerical studies suggest that super-Eddington accretion can also be radiatively efficient for optimal conditions (Y.-F. Jiang et al. 2014; J. C. McKinney et al. 2015). Furthermore, IMBHs may not only be harbored in centers of small galaxies but also be hosted in globular clusters (GCs). It has also been shown that during galaxy mergers the satellite galaxies that potentially host IMBHs take a long time to merge with the central SMBH (H. Pfister et al. 2021). Therefore, some fraction of IMBH TDEs are likely to be found off-center in massive galaxies. Last but not least, IMBHs can disrupt compact stars such as white dwarfs (M. MacLeod et al. 2016; J. Law-Smith et al. 2017), which also produces relatively strong gravitational waves. This motivates the prospect of conducting multimessenger studies of TDEs (M. Eracleous et al. 2019; H. Pfister et al. 2022).

While it has been speculated that a large fraction of the TDEs can be produced from IMBHs, the TDE rate calculation in this regime has been severely lacking. Only very recently have M. Polkas et al. (2024) calculated the time-dependent rate of TDEs including the contribution from the IMBH population from the centers of galaxies, by analyzing the results from a semianalytical galaxy formation and evolution model. They report that as $M_{\rm BH}$ increases, the TDE rate first increases until $M_{\rm BH} \sim 10^{5.5}$ and then flattens before turning over to decrease with $M_{\rm BH}$ at higher masses. However, in their model, the NSC component, which impacts the TDE rates significantly, was implemented based on a simple phenomenological model. On the other hand, for IMBHs in GCs, E. Ramirez-Ruiz & S. Rosswog (2009) estimated their TDE volumetric rates to be around $4 \times 10^{-6}$ yr$^{-1}$ Mpc$^{-3}$. Recently, V. L. Tang et al. (2024) further estimated that the TDE rate from an individual GC lies between $10^{-8}$ and $10^{-5}$ yr$^{-1}$, with a dependence on the GC stellar mass.

We aim to fill the gaps in these estimates by calculating the IMBH TDE rate from a different perspective. In this work, we adopt realistic stellar profiles for galaxies and stellar clusters hosting IMBHs, which are obtained from recent state-of-the-art observations, and conduct loss cone dynamics calculations to obtain the TDE rates. We note that these IMBH-hosting galaxies and clusters can only be resolved at very low redshifts. After calculating the TDE rates from individual IMBHs, we apply the BH mass function (BHMF) and occupation fraction, also obtained from observations, to calculate the volumetric IMBH TDE rates.

Our Letter is structured as follows: We provide a description of the methodology in Section 2, where we introduce the observed samples of IMBH-hosting galaxies and stellar clusters, as well as the loss cone dynamics calculation. We report the computed IMBH TDE rates and compare them to SMBH TDE rates in Section 3.1, and we calculate the contribution from the pinhole regime and the diffusive regime to the rates in Section 3.2. We analyze the dependence of TDE rates and pinhole fraction on the stellar profile in Section 3.3. Next, we calculate the volumetric IMBH and SMBH TDE rates and compare them to observed rates in Section 3.4. The distribution of the penetration parameter $\beta$ in IMBH and SMBH TDEs is computed and presented in Section 3.5. We then investigate the behavior of the TDE rate as a function of BH mass in detail in Section 3.6. Lastly, we draw a summary and further discuss our results in Section 4.

## 2. Methodology

### 2.1. IMBH Sample

We collate a sample of 22 observed IMBHs for our TDE rate calculation. These IMBHs are all nearby with dynamically measured masses. This ensures that the stellar profiles close to the IMBHs are also well constrained, which permits accurate TDE rate calculations.

For constructing this IMBH sample, we start from the IMBHs listed in Tables 2–4 in J. E. Greene et al. (2020) and only select those with dynamical BH mass measurements. We also add a recently observed IMBH from GC B023-G78 (R. Pechetti et al. 2022).

Among these 22 IMBHs, 14 reside in the centers of their host galaxies (hereafter referred to as IMBH-GN, where "GN" stands for "galactic nucleus"); these are listed in Table A1 of Appendix A. The other eight IMBH candidates reside in GCs that are off-nuclear (hereafter referred to as IMBH-ON, where "ON" stands for "off-nucleus"); these are listed in Table A2 of Appendix A.

The BH mass is a parameter that impacts the TDE rate and demographics calculation. Several IMBH-GNs in our list from N. Neumayer & C. J. Walcher (2012) are reported with both a best mass estimate ($M_{\rm BH,best}$) and a maximum mass constraint ($M_{\rm BH,max}$), while the others only have a single mass estimate.

For IMBH-ONs, their mass measurements are less accurate, and sometimes for the same IMBH the $M_{\rm BH,best}$ measurement from one published source differs from and can be larger than the $M_{\rm BH,max}$ constraint from another reported measurement. Here we adopt the largest value of the reported $M_{\rm BH}$ values for IMBH-ONs and consider them all as $M_{\rm BH,max}$.

We then further split the IMBH sample into two groups based on the quality of the BH mass measurements: (1) quality measurements—IMBH-GNs with $M_{\rm BH,best}$ measurements, except those from N. Neumayer & C. J. Walcher (2012); and (2) tentative measurements—IMBHs with $M_{\rm BH,max}$ constraints only and the IMBH-GNs from N. Neumayer & C. J. Walcher (2012). We consider the $M_{\rm BH,best}$ reported by N. Neumayer & C. J. Walcher (2012) as tentative measurements since their calculation is based on a single integrated velocity dispersion that is dominated by the stellar mass at large radii (N. Neumayer et al. 2020). Tentative measurements are weighted down by an arbitrary factor of 50% when obtaining several fits in later sections. There is also one special case of NGC 4395, for which the IMBH mass has been measured using both dynamical measurement (e.g., M. den Brok et al. 2015) and reverberation (e.g., J.-H. Woo et al. 2019; H. Cho et al. 2021) techniques, and the two $M_{\rm BH}$ values have an order-of-magnitude discrepancy. For this IMBH, we take the geometric mean of these two $M_{\rm BH}$ values for the TDE rate calculation.

### 2.2. Stellar Profiles of the Galaxies and Clusters Hosting IMBHs

The stellar profiles of the galaxies or clusters hosting the IMBHs in our sample have all been previously modeled using either the Sérsic profile (A. J. Barth et al. 2009; N. Neumayer & C. J. Walcher 2012; M. den Brok et al. 2015; B. Kızıltan et al. 2017; D. D. Nguyen et al. 2019; T. A. Davis et al. 2020) or the King profile (J. Gerssen et al. 2002; E. Noyola et al. 2010; N. Lützgendorf et al. 2013, 2015; H. Baumgardt 2017; B. B. P. Perera et al. 2017; R. Pechetti et al. 2022). For the galaxies hosting IMBH-GNs, we adopt their stellar profiles fitted to Sérsic models, which is described further in Section 2.2.1. However, the stellar profiles of the clusters hosting IMBH-ONs have only been fitted using the King model in the literature, and this in turn is described in Section 2.2.2. We list the fitted parameters for the IMBH-GC and IMBH-ON sources in the collated sample in Appendix A, Tables A1 and A2, respectively.

Low-mass galaxies are often observed to host NSCs (N. Neumayer et al. 2020). Therefore, one can see that some galaxies in Appendix A Table A1 can have more than one component describing their stellar profile. Moreover, we note that the eight galaxies from N. Neumayer & C. J. Walcher (2012) have measurements of the stellar profiles of only their NSCs but not of their bulges. This, however, does not affect our TDE rate calculation, since NSCs are significantly denser than bulges and therefore are expected to dominate the TDE rates (H. Pfister et al. 2020). Similarly, B023-G78 was also modeled with an inner King profile and an outer Sérsic profile. For the same reason, we have only considered the fitted inner King profile for the TDE rate calculations.

#### 2.2.1. Sérsic Profile Fits for Galaxies

The Sérsic profile, which is commonly used to model the surface brightness of a galaxy, is described by three fundamental parameters: the Sérsic index $n$, the effective radius $R_{\rm eff}$, and a normalization parameter. The stellar density profile can then be approximated by (P. Prugniel & F. Simien 1997)

$$\rho(r) = \rho_0 \left(\frac{r}{R_{\rm eff}}\right)^{-p} e^{-b(r/R_{\rm eff})^{1/n}} \quad (1)$$

$$p = 1 - \frac{0.6097}{n} + \frac{0.05563}{n^2} \quad (2)$$

$$b = 2n - \frac{1}{3} + \frac{0.009876}{n}. \quad (3)$$

Here one can see that the stellar density profile is described with three independent parameters: (1) the effective radius $R_{\rm eff}$; (2) the Sérsic inner slope $p$, which is a function of the Sérsic index $n$ and describes the steepness of the stellar density slope within $R_{\rm eff}$; and (3) the central stellar density $\rho_0$, which is given by

$$\rho_0 = \frac{M_\star}{4\pi R_{\rm eff}^3} \frac{b^{n(3-p)}}{n\, \gamma_{\rm E}(n(3-p))}. \quad (4)$$

Here $M_\star$ is the total stellar mass of the galaxy bulge/NSC or the cluster, and $\gamma_{\rm E}$ is the Euler Gamma function:

$$\gamma_{\rm E}(x) = \int_0^\infty t^{x-1} e^{-t} dt. \quad (5)$$

For the TDE rate calculation, the stellar density profile needs to be converted into the stellar distribution function (DF), which is the number density of stars as a function of the orbital specific energy $E$, using the Eddington formula:

$$f(E) = \frac{1}{\sqrt{8}\,\pi^2 m} \frac{d}{dE} \int_0^E \frac{d\rho}{d\psi} \frac{d\psi}{\sqrt{E-\psi}}. \quad (6)$$

There is no analytic solution for the above equation, however, for the stellar profiles described by the Sérsic profile. Therefore, we compute the DF numerically through PHASE-FLOW (E. Vasiliev 2017), the details of which are described in Section 2.5. We note that mathematically $p$ has to be $> 0.5$ for the $f(E)$ to be positive. Hence, for systems with $p < 0.5$ we manually replace $p$ to be 0.5 when computing the DF.

#### 2.2.2. King Profile Fits for GCs

The King model is often used to describe GCs, and the density profile is given by

$$\rho(\Psi) = \rho_1 \left[ e^{\Psi/\sigma^2} {\rm erf}\left(\frac{\sqrt{\Psi}}{\sigma}\right) - \sqrt{\frac{4\Psi}{\pi\sigma^2}}\left(1 + \frac{2\Psi}{3\sigma^2}\right) \right], \quad (7)$$

where $\rho_1$ is a scaling parameter, $\sigma$ is the one-dimensional velocity dispersion, and $\Psi$ is the relative gravitational potential between $r$ and the tidal radius $r_t$:

$$\Psi(r) = \Phi(r_t) - \Phi(r). \quad (8)$$

The King model is parameterized by the king radius $r_s$ and the dimensionless potential $W_0 \equiv \Psi(0)/\sigma^2$ (I. R. King 1966). We note that one could also parameterize the King model using the concentration parameter $C$ instead of $W_0$. For the IMBH-ON in GCs with only $C$ provided in the literature, we use the one-to-one correspondence between $W_0$ and $C$ to obtain the value of $W_0$ (J. J. Binney 2011).

Since the King profile is basically a lowered isothermal model, its DF is similar to that of an isothermal sphere at small radii, but it is truncated at larger radii to have a finite mass. The DF for this profile has an analytic form (J. J. Binney 2011):

$$f(E) = \begin{cases} \frac{\rho_1}{(2\pi\sigma^2)^{3/2}}(e^{E/\sigma^2} - 1) & \text{when } E > 0 \\ 0 & \text{when } E < 0 \end{cases}. \quad (9)$$

### 2.3. SMBH Sample and Host Galaxy Profiles

For comparison with IMBHs, we also utilize 37 SMBHs from nearby galaxies with their dynamically determined BH masses $M_{BH}$. These data are taken from I. Trujillo et al. (2004), D. D. Nguyen et al. (2018), and B. L. Davis et al. (2019), who also provide Sérsic profiles for the host galaxies. The TDE rates in these galaxies have been previously calculated by H. Pfister et al. (2020). We further select the BHs with masses between $10^6$ and $10^8$ $M_\odot$ for our SMBH sample. A few of these galaxies with lower masses contain one or more NSC components. We follow H. Pfister et al. (2020) and adopt their NSC fit of the Milky Way, along with two other galaxies from R. Pechetti et al. (2020). The parameters of the SMBHs and their hosting galaxies are listed in Table A3 of Appendix A.

### 2.4. Loss Cone Dynamics

We provide a quick overview of the loss cone, but we refer readers to L. E. Strubbe (2011) and N. C. Stone et al. (2020) for a more in-depth review. A star with mass $m_\star$ and radius $r_\star$ will be tidally disrupted by an MBH of mass $M_{BH}$ if it gets closer than the tidal disruption radius

$$r_t \equiv (M_{BH}/m_\star)^{1/3} r_\star. \quad (10)$$

Specific orbital angular momentum of the star at $r_t$ can be written as $L_{lc} = \sqrt{2GM_{BH}r_t}$. Hence, it can also be said that the star will get disrupted if its angular momentum $L < L_{lc}$. This phase-space region is commonly known as the loss cone. We can further define a variable

$$R_{lc}(E) = \frac{L_{lc}^2}{L_{circ}^2(E)} = \frac{2GM_{BH}r_t}{L_{circ}^2(E)}, \quad (11)$$

where $L_{circ}(E)$ is the specific angular momentum of a circular orbit.

Based on this variable and orbital period $P(E)$, we can define the loss cone filling factor such that

$$q(E) = \frac{P(E)\bar{u}(E)}{R_{lc}(E)}, \quad (12)$$

where $\bar{u}(E)$ is the orbital-averaged diffusion coefficient.

When $q \ll 1$, a star needs multiple orbital periods to be scattered into the loss cone, and hence its orbit typically has a low $\beta$ value. This constitutes the diffusive regime. On the opposite end, when $q \gg 1$, stars can scatter in and out of the loss cone within a single orbital period, and therefore their orbits can have any $\beta$, which enhances the probability of producing TDEs on deeply plunging orbits. This region is known as the pinhole regime.

### 2.5. The TDE Rate Calculation

We mainly follow the approach of H. Pfister et al. (2020) to calculate TDE rates, which we briefly summarize below.

We start by utilizing PHASEFLOW (E. Vasiliev 2017), which takes the observed stellar profiles described by the Sérsic or King model and returns a corresponding numerical DF. To be more specific, the following parameters are computed by PHASEFLOW:

1. $\bar{f}(E)$, the averaged DF over angular momentum;
2. $\bar{u}(E)$, the orbital-averaged diffusion coefficient;
3. $L_{circ}(E)$, the angular momentum of a circular orbit of binding energy $E$; and
4. $P(E)$, the orbital period.

We then follow the prescriptions described in L. E. Strubbe (2011) to compute the TDE rates. The differential TDE rate can be computed using the following equation:

$$\frac{d^2\Gamma}{dE d\ln\beta} = -8\pi^2 G M_{BH} \frac{r_t}{\beta} \mathcal{G}(q(E), \beta), \quad (13)$$

where $\beta \equiv r_t/r_p$ is the penetration parameter defined by the ratio of the tidal disruption radius $r_t$ and the pericenter radius $r_p$. The upper and lower limits of $\beta$ depend on parameters such as the BH spin (M. Kesden 2012) and stellar models (J. Guillochon & E. Ramirez-Ruiz 2013). Here, for simplicity, we set $\beta \in [1, r_t/r_{sch}]$, where the upper limit corresponds to the stellar orbit with its pericenter at the Schwarzschild radius of the BH $r_{sch}$. The choice of $\beta_{max}$ can indeed impact the TDE rate calculated, especially when $M_{BH}$ is very large, given that the range of $\beta$ is narrow. However, for IMBH TDEs, orbits with $\beta \sim \beta_{max}$ are very steep encounters, which contributes very little to the total TDE rate as we will show. Therefore, the TDE rate is not very sensitive to $\beta_{max}$.

Finally, $\mathcal{G}$ is given by the following equations:

$$\mathcal{G}(q, \beta) = \frac{\bar{f}(E)}{1 + q^{-1}\xi(q)\ln(1/R_{lc})} \times \left[1 - 2\sum_{m=1}^{\infty} \frac{e^{-\alpha_m^2 q/4}}{\alpha_m} \frac{J_0(\alpha_m \beta^{-1/2})}{J_1(\alpha_m)}\right] \quad (14)$$

$$\xi(q) = 1 - 4\sum_{m=1}^{\infty} \frac{e^{-\alpha_m^2 q/4}}{\alpha_m^2}, \quad (15)$$

where $J_i$ is the Bessel function of $i$th order and $a_m$ is the $m$th zero of $J_0$.

### 2.6. Stellar Mass Function for Disrupted Stars

In this work, we take the mass distribution of stars $\phi(m)$ explicitly into account instead of using a monochromatic stellar mass as done by previous groups. Following previous work (H. Pfister et al. 2022; T. H. T. Wong et al. 2022), we use the Kroupa initial mass function (IMF) with the stellar mass $m \in [0.08, 10]$, which has the form (P. Kroupa 2001)

$$\phi(m) \propto \begin{cases} m^{-1.3} & \text{for } m < 0.5 \\ m^{-2.3} & \text{for } m > 0.5 \end{cases}. \quad (16)$$

For a distribution of stellar masses, we have the following total differential TDE rate:

$$\frac{d^2\Gamma}{dE d\ln\beta} = \int \frac{d^2\Gamma(m)}{dE d\ln\beta} m \, \phi(m) \, d\ln m. \quad (17)$$

For computing the above, we use $f = f_{mono}/\langle m \rangle$ and $\bar{u} = \bar{u}_{mono}\langle m^2 \rangle / \langle m \rangle$, where $f_{mono}$ and $\mu_{mono}$ are the DF and diffusion coefficient of a monochromatic stellar mass distribution, respectively, as described in Section 2.5 (see the derivation of these two equations in Appendix B); $\langle m \rangle = \int m \, \phi(m) \, dm$ is the averaged mass of stars; and $\langle m^2 \rangle = \int m^2 \phi(m)\, dm$ is the mean square of the stellar mass distribution. As a result, the overall TDE rate can vary by a few times when using different cutoffs for the stellar mass distributions (N. C. Stone & B. D. Metzger 2016).

## 3. Results

### 3.1. TDE Rates from IMBHs and SMBHs

We calculate TDE rates following the method outlined in Section 2.5. The TDE rates from IMBH-GNs and IMBH-ONs compared to those computed from SMBHs are plotted in Figure 1.

In Figure 1, panel (a) shows the TDE rates $\Gamma$ from individual galaxies or clusters plotted against $M_{BH}$. For the IMBH population studied here, we denote the rates computed with $M_{BH,best}$ and $M_{BH,max}$ using filled and open symbols, respectively, and different colors for the IMBH-GN TDEs (red), IMBH-ON TDEs (blue), and all SMBH TDEs (green). For galaxies that are fitted with more than one component, the TDE rates from these components are summed together, although we note that the innermost NSC dominates the rate calculation. We also calculate the uncertainty of $\Gamma$ based on the uncertainty of $M_{BH}$ measurements mentioned in Appendix A.

Overall, the TDE rates from IMBHs range between $\sim 10^{-8}$ and $10^{-4}$ galaxy$^{-1}$ yr$^{-1}$, which is comparable to those from SMBHs. One can see the strikingly opposite trends of $\Gamma$ versus $M_{BH}$ in the IMBH and SMBH regimes. For SMBHs, $\Gamma$ on average decreases as $M_{BH}$ increases, which is qualitatively consistent with earlier results from J. Wang & D. Merritt (2004), N. C. Stone & B. D. Metzger (2016), and H. Pfister et al. (2020). However, we find that for the population of IMBHs $\Gamma$ generally increases with increasing $M_{BH}$. On average, the TDE rate peaks around $M_{BH} \sim 10^6 \, M_\odot$. This turnover in TDE rate as a function of $M_{BH}$ is qualitatively consistent with the results by M. Polkas et al. (2024). They suspect that this behavior is a consequence of the remarkably different stellar profiles of NSCs and bulges. We conduct a further investigation on the cause of this turnover in Section 3.6.

We next calculate the TDE rate $\Gamma$ as a function of $M_{BH}$ in Figure 1(b). Based on the individual galaxy/cluster TDE rate shown in panel (a), we make histograms of averaged TDE rates in each $M_{BH}$ bin, which are shown as the blue shaded regions in panel (b). For this calculation the samples with tentative $M_{BH}$ measurements are down-weighted by half. It is clearly seen that the averaged TDE rate in an $M_{BH}$ bin is dominated by galaxies with higher TDE rates. Then, we fit the histograms to obtain the $\Gamma$–$M_{BH}$ with a broken power-law function, with the turnover manually placed at $M_{BH} = 10^6 \, M_\odot$. The best-fit functions plotted (using blue lines) are

$$\Gamma = \begin{cases} 1.2 \times 10^{-4} \left(\dfrac{M_{BH}}{10^6 \, M_\odot}\right)^{1.2} \text{yr}^{-1} & \text{for} \quad M_{BH} \leqslant 10^6 \, M_\odot \\ 1.2 \times 10^{-4} \left(\dfrac{M_{BH}}{10^6 \, M_\odot}\right)^{-1.2} \text{yr}^{-1} & \text{for} \quad M_{BH} > 10^6 \, M_\odot \end{cases}.$$

(18)

We then estimate the averaged TDE rates around IMBH-GNs, IMBH-ONs, and SMBHs separately. Based on the mass ranges of these three BH populations, we calculate the maximum and minimum values within the mass range using the best-fit function (Equation (18)) and then take a linear average of the two. The BH mass ranges are chosen to be $M_{BH} \in [10^2, 10^6]$ for IMBH-GNs (with the lower limit placed around the lowest observed IMBH-GN mass in our sample), $M_{BH} \in [2 \times 10^3, 9.1 \times 10^4]$ for IMBH-ONs (with both limits based on the range of the observed IMBH-ON masses), and $M_{BH} \in [10^6, 10^8]$ for SMBHs (with the upper limit placed around the Hill mass of an SMBH, which can still disrupt a solar-type star outside its event horizon). The averaged TDE rates are then calculated to be $\Gamma_{avg} \approx 1.3 \times 10^{-4}$ galaxy$^{-1}$ yr$^{-1}$ for IMBH-GNs, $\Gamma_{avg} \approx 6.9 \times 10^{-6}$ galaxy$^{-1}$ yr$^{-1}$ for IMBH-ONs, and $\Gamma_{avg} \approx 1.3 \times 10^{-4}$ galaxy$^{-1}$ yr$^{-1}$ for SMBHs. We note that these estimates are just used to give an indication that the TDE rates (per galaxy) for IMBH-GNs and SMBHs are comparable, and that for IMBH-ONs should be lower. The actual observed TDE rate for each BH population should depend on the distribution of the BH masses, which will be discussed in Section 3.4.

We now plot the TDE rates for IMBHs and SMBHs against the total stellar mass $M_\star$ of the host galaxy or cluster in Figure 1(c). We see that there exists a clear gap in the stellar mass between clusters hosting IMBH-ONs and the others. Some of the galaxies hosting IMBH-GNs are considered dwarf galaxies, with total stellar masses $M_\ast < 3 \times 10^9 \, M_\odot$. However, surprisingly, most galaxies hosting IMBH-GNs have stellar masses that are larger; in fact, they are only slightly smaller than those of galaxies hosting SMBHs. This indicates that the $M_{BH}$ scaling relations for SMBHs likely break down for IMBHs. We note that this is consistent with previous findings (J. E. Greene & L. C. Ho 2006; I. Martìn-Navarro & M. Mezcua 2018). Additionally, the majority of our sample of galaxies hosting IMBHs contain dense NSCs. Therefore, we also plot the TDE rate against the mass of the innermost stellar component in Figure 1(d). Here we denote the innermost stellar component of a galaxy using a star symbol if it is an NSC, or we continue using a circle if it is a bulge. All galaxies hosting IMBH-GNs have NSCs in the center, whereas most galaxies hosting SMBHs only have bulges. However, it is not straightforward to infer the IMBH TDE rates using either the total stellar mass or the stellar mass of the innermost stellar component.

Last but not least, as one can notice from Figure 1(a), the IMBH mass measurements generally bear a larger degree of uncertainty compared to those of SMBHs, which leads to higher uncertainty in the IMBH TDE rate estimates. In addition, we note that the Sérsic model is limited by its shallow stellar density slope ($p < 1$), which can result in underestimating the TDE rate. Furthermore, the choice of the stellar profile model can also lead to discrepancies in the TDE rate estimates. While we have used the Sérsic stellar density profiles for calculating the TDE rates from IMBH-GNs and SMBHs, the clusters hosting IMBH-ONs are only described with the King model in the literature. However, the effect of this discrepancy tends to be small, and the difference in rate estimates is usually within 0.5 dex (N. C. Stone & B. D. Metzger 2016). This implies that, to first order, our results and inferred trends are trustworthy and reliable.

### 3.2. The TDE Pinhole Fraction

The TDE pinhole fraction $f$ denotes the fraction of TDEs in the pinhole regime:

$$f \equiv \frac{\Gamma_{q>1}}{\Gamma_{tot}}, \quad (19)$$

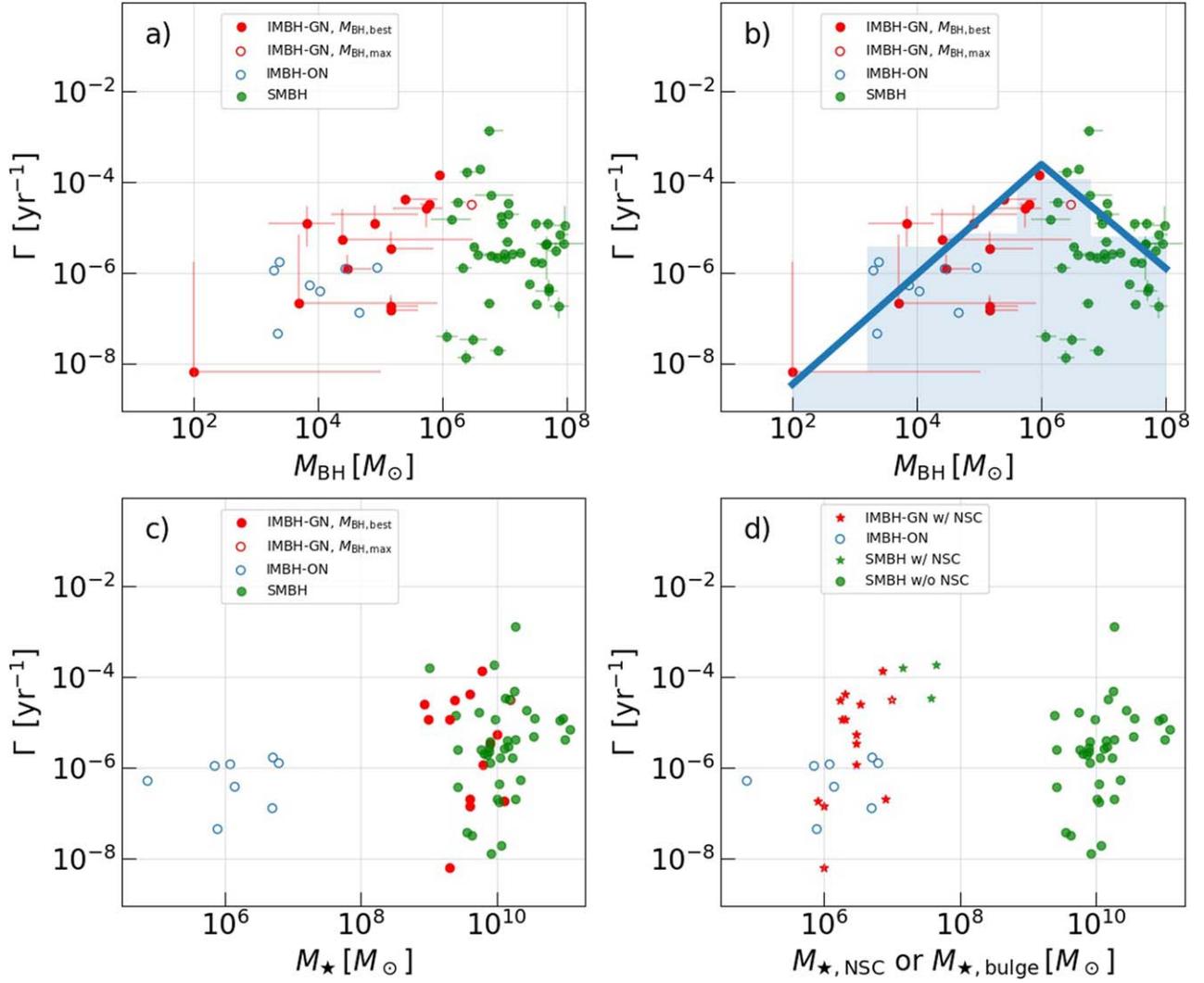

**Figure 1.** TDE rate $\Gamma$ for each galaxy or cluster in our sample. Three different populations of MBHs are shown, denoted with different colors: all IMBH-GNs in red, all IMBH-GCs in blue, and all SMBHs in green. Panel (a) shows $\Gamma$ plotted against $M_{\rm BH}$. We adopt $M_{\rm BH,best}$ (filled symbols) for an MBH if a best mass measurement exists; otherwise, we use $M_{\rm BH,max}$ (open symbols) if only a maximum mass measurement exists. Panel (b) is the same as panel (a) but includes additional histograms (blue shaded regions) showing the averaged TDE rate distribution and a broken power-law fit (blue line) for the averaged $\Gamma$–$M_{\rm BH}$ relation (Equation (18)). One can see that $\Gamma$ increases with $M_{\rm BH}$ in the IMBH regime but decreases with $M_{\rm BH}$ in the SMBH regime. Panel (c) shows $\Gamma$ plotted against the total stellar mass of a galaxy or cluster $M_\star$. Panel (d) shows $\Gamma$ plotted against the mass of the innermost stellar component when a galaxy can be fitted with multiple components. Galaxies with an NSC as the innermost component are represented using star symbols instead of circles.

where $\Gamma_{q>1}$ denotes the rate of TDEs with $q > 1$ and $\Gamma_{\rm tot}$ denotes the total TDE rate. We calculate $f$ for all three populations studied here, IMBH-GNs, IMBH-ONs, and SMBHs, and plot $f$ versus $M_{\rm BH}$ and $f$ versus $M_\star$ in Figure 2. For SMBHs we see that on average $f$ decreases as $M_{\rm BH}$ increases. This is consistent with the result reported earlier in N. C. Stone & B. D. Metzger (2016). Interestingly, for both IMBH-GNs and IMBH-ONs, $f$ generally stays close to unity regardless of $M_{\rm BH}$. This means that IMBH TDEs mostly reside in the pinhole regime. Even for SMBHs, TDEs only start to be mainly in the diffusive regime when $M_{\rm BH}$ becomes large enough. A similar trend with $M_\star$ is also found and shown.

The relation between $f$ and $M_{\rm BH}$ is determined by multiple physical factors. If one keeps the stellar profile the same but varies $M_{\rm BH}$, it is found that a larger $M_{\rm BH}$ will lead to a lower $f$. There are two reasons for this. First, increasing $M_{\rm BH}$ moves the tidal disruption radius $r_T$ closer to the BH event horizon. Second, the loss cone size also increases with $M_{\rm BH}$. Both these effects cohere to increase the fraction of TDEs in the diffusive regime. This is explained in more detail in Appendix C and illustrated with the case of NGC 1042. Furthermore, $f$ depends on the stellar density profile, which has an implicit dependence on $M_{\rm BH}$, which we address in the next section.

We provide the following fitting functions for $f$ versus $M_{\rm BH}$ and $f$ versus $M_\star$:

$$f(M_{\rm BH}) = 1 - \left(\frac{M_{\rm BH}}{10^8 \, M_\odot}\right)^{2.5} \tag{20}$$

$$f(M_\star) = 1 - \left(\frac{M_\star}{10^{11} \, M_\odot}\right)^{1.5}. \tag{21}$$

### 3.3. Impact of the Stellar Profile of the Host Galaxy or Cluster

In this subsection, we investigate how the stellar density profiles of IMBH/SMBH host galaxies or clusters affect their TDE rates and pinhole fractions. As a reminder, the stellar profiles of galaxies hosting SMBHs and IMBH-GNs in our sample are modeled using the Sérsic profile, which is described by three fundamental parameters (Equations (1)–(4)): (1) the Sérsic inner slope $p$, (2) the stellar central density $\rho_0$, and (3) the effective radius $R_{\rm eff}$. Clusters hosting IMBH-ONs in our sample are only modeled using the King profile.

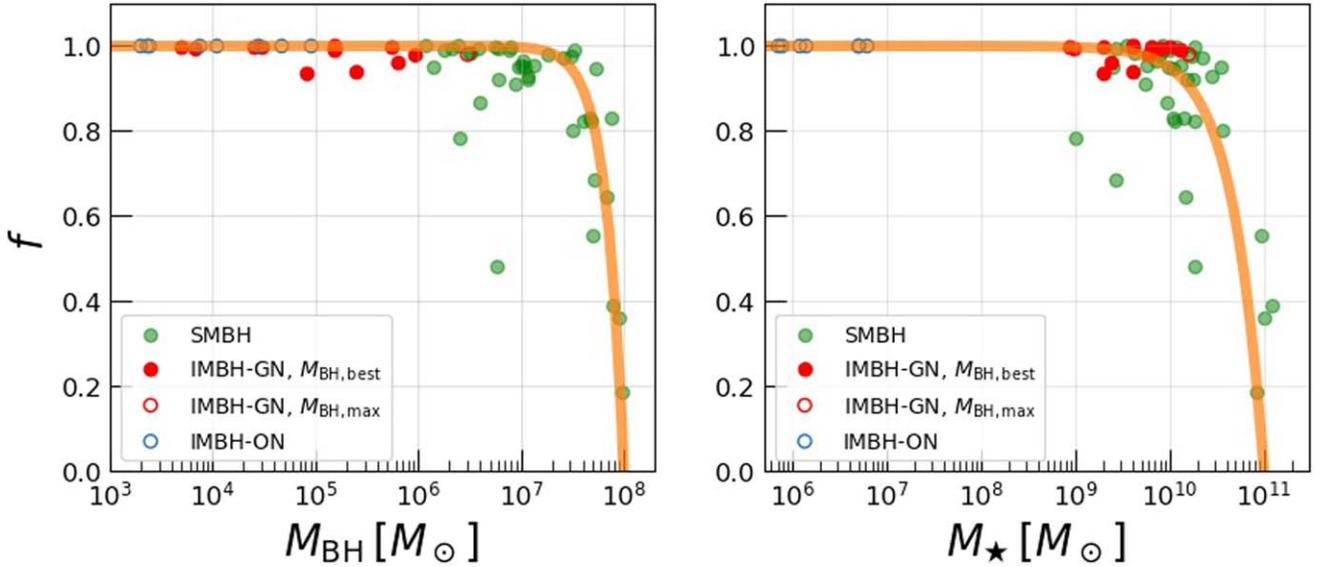

**Figure 2.** Pinhole fraction $f$ plotted against $M_{BH}$ and $M_\star$. The symbols and colors are the same as in Figure 1. The orange curves represent the fitting functions (Equations (20) and (21)).

We plot the stellar density profiles $\rho(r)$ of all galaxies or clusters hosting IMBH-GNs and IMBH-ONs in Figures 3(a) and (b), respectively, using solid curves. For comparison, we also calculate the mean and the standard deviation $\sigma$ of $\rho(r)$ for all SMBH-hosting galaxies, and we show the $1\sigma$ range of the mean value using the blue shaded region. In general, the galaxies hosting IMBH-GNs appear to have higher stellar central densities, similar inner density slopes, and smaller effective sizes compared to the galaxies hosting SMBHs. On the other hand, the clusters hosting IMBH-ONs appear to have similar stellar central densities, steeper stellar density slopes, and smaller sizes compared to galaxies hosting SMBHs.

We next investigate the implicit relations between the BH mass and host galaxy/cluster stellar profiles in our sample. We plot $M_{BH}$ versus $p$, $M_{BH}$ versus $\rho_0$, and $M_{BH}$ versus $R_{\rm eff}$ in Figures 4(a)–(c), respectively, for IMBH-GNs and SMBHs in our sample. We exclude IMBH-ONs from these plots, as the $x$-axis parameters are relevant only for the Sérsic model. For galaxies with both bulge and NSC components, we only use the parameters of the innermost NSC component and indicate that using star symbols. The stellar profile parameter comparisons between IMBH-GN and SMBH hosts are consistent with Figure 3(a). IMBH-GN host galaxies have higher central stellar densities since they all have NSC components. Their stellar density slopes are on average similar to those of SMBH hosts. For both populations, there exists a trend that on average $p$ increases with $M_{BH}$. This is consistent with previous findings by G. Savorgnan et al. (2013) but different from the trend reported by N. C. Stone & B. D. Metzger (2016), who have shown a transition from cuspy to core galaxies as the $M_{BH}$ increases. This discrepancy could be attributed to differences in the sample compositions. While our sample consists of IMBHs and SMBHs with masses up to $10^8\,M_\odot$, the sample used by N. C. Stone & B. D. Metzger (2016) consists of SMBHs with masses up to $\sim 10^{10}\,M_\odot$. Furthermore, the $R_{\rm eff}$ of the innermost NSCs in both IMBH-GN and SMBH hosts are generally quite compact, with $R_{\rm eff}$ ranging mostly between $10^0$ and $10^1$ pc. The NSC size does not appear to strongly correlate with $M_{BH}$. On the other hand, the bulges of galaxies hosting SMBHs have larger sizes ($R_{\rm eff} > 10^2$ pc), and the bulge size generally increases with $M_{BH}$.

We also plot the correlation between the TDE rate from each galaxy and the respective stellar profile parameters, $\Gamma$ versus $p$, $\Gamma$ versus $\rho_0$, and $\Gamma$ versus $R_{\rm eff}$, in Figures 4(d)–(f). Figure 4(d) shows that $\Gamma$ generally increases with the inner slope $p$ for both populations, although a large scatter is seen. This can be interpreted as follows: a steeper $p$ puts more stars closer to the MBH, increasing the probability of scattering stars into the loss cone and hence enhancing the TDE rate. The $\Gamma$ versus $p$ relations for IMBH-GNs and SMBHs can be fit with the following power-law functions (excluding galaxies with $p < 0.5$):

$$\Gamma = \begin{cases} 8.7 \times 10^{-7}\,10^{5.9(p-0.5)} & {\rm yr}^{-1}\quad \text{for IMBH-GNs} \\ 7.1 \times 10^{-7}\,10^{3.5(p-0.5)} & {\rm yr}^{-1}\quad \text{for SMBHs} \end{cases}.$$
(22)

Figure 4(e) shows a clear positive correlation between $\Gamma$ and $\rho_0$ for each population. This is not surprising—it is expected that the TDE rate should scale with the overall stellar density, as there are more stars available for disruption. Once again, we see very similar trends in the $\Gamma$ versus $\rho_0$ relations for IMBHS-GNs and SMBHs, albeit with a small offset. This offset is a result of the difference in $R_{\rm eff}$ between the two MBH populations. We further show the relations between $\Gamma$ and two other densities in Appendix D. The best-fit power-law relations for $\Gamma$ versus $\rho_0$ are

$$\Gamma = \begin{cases} 2.5 \times 10^{-10}\,\rho_0^{0.93} & {\rm yr}^{-1}\ \text{for IMBH-GNs} \\ 1.2 \times 10^{-7}\,\rho_0^{0.65} & {\rm yr}^{-1}\ \text{for SMBHs} \end{cases}. \quad (23)$$

On the other hand, Figure 4(f) shows that $R_{\rm eff}$ does not affect $\Gamma$.

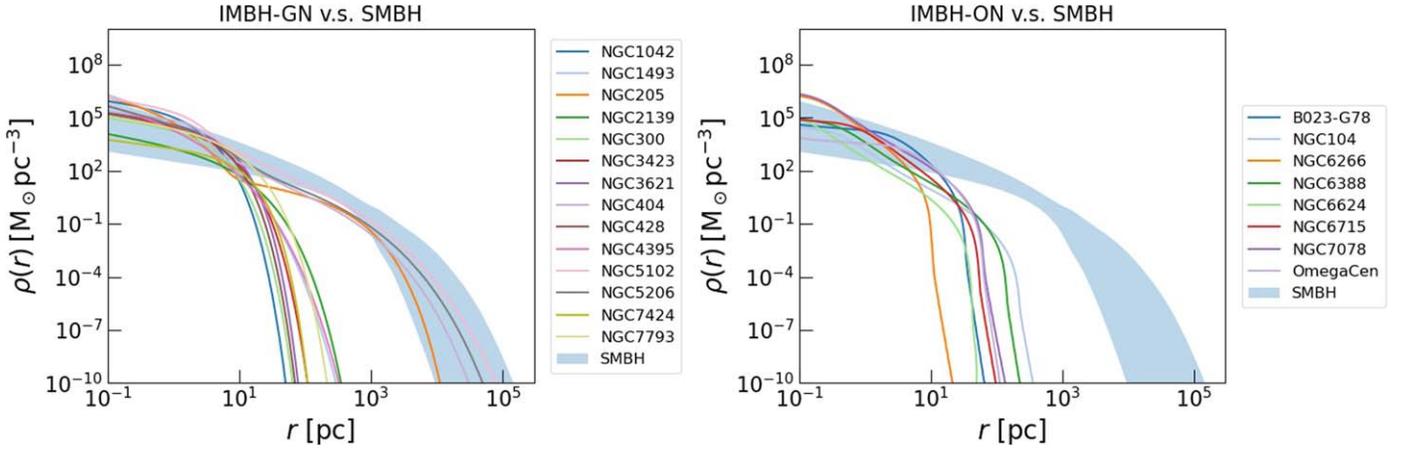

**Figure 3.** The stellar density profiles of galaxies and clusters hosting IMBH-GNs (left panel) and IMBH-ONs (right panel). The shaded region shows the 1σ range of the stellar density profiles of the galaxies hosting SMBHs for comparison.

Lastly, we show how the TDE pinhole fraction $f$ in each galaxy depends on the stellar density parameters, $f$ versus $p$, $f$ versus $\rho_0$, and $f$ versus $R_{\rm eff}$, in Figures 4(g)–(i), respectively. Here the SMBH population is further split into two groups according to their masses: $10^6$–$10^7\ M_\odot$ (turquoise circles) and $>10^7\ M_\odot$ (orange circles). One can see that $p$ is the most important parameter determining $f$: $f$ typically decreases with increasing $p$. When $p$ is below $\sim 0.7$, $f$ is approximately unity. Once $p \gtrsim 0.7$, $f$ gradually decreases toward 0. To understand how $p$ impacts $f$, one can imagine a galaxy with a very flat inner density slope (i.e., with very small $p$). In such a hypothetical scenario, more stars are farther from the MBH, so more energetic scatterings will be required for stars to enter the loss cone, i.e., more TDEs will be disrupted in the pinhole regime. Interestingly, we also find that all galaxies with low $f$ and high $p$ values host heavy SMBHs with $M_{\rm BH} > 10^7\ M_\odot$. Since stars are more likely disrupted by lighter MBHs, this means that the majority of TDEs should occur in the pinhole regime.

The dependence of $f$ on $\rho_0$ or $R_{\rm eff}$ is subtler. From Figure 4(h) no obvious trend is seen when looking at all the galaxy populations in concert. This is expected since $\rho_0$ is only a scaling factor for the stellar density, which does not affect $f$. However, after decomposing the galaxy populations into three categories, IMBHs, lighter SMBHs, and heavier SMBHs, it is seen that $f$ decreases with increasing $\rho_0$ for each galaxy population. This is because when we fix the range of $M_{\rm BH}$, we somewhat confine the range of $M_\star$ to wherein a larger $\rho_0$ would result in a larger $p$. As for Figure 4(i), we do not notice any obvious, direct trend between $f$ and $R_{\rm eff}$.

In summary, there exist positive correlations between the TDE rate and two stellar density profile parameters, $\rho_0$ and $p$, separately for the IMBH-GN and SMBH populations. On the other hand, the TDE pinhole fraction $f$ primarily depends on $p$. However, there exist implicit relationships between these parameters when we consider a specific MBH mass range. For example, almost all host IMBH-GNs and SMBHs with $M_{\rm BH} < 10^7\ M_\odot$ in our sample have stellar profiles with $p \lesssim 0.7$, which means that the TDEs from these systems mostly occur in the pinhole regime.

### 3.4. Volumetric Rates of IMBH and SMBH TDEs

In this section we compute the volumetric TDE rates for IMBHs and SMBHs. The volumetric rate $\dot{N}$ is computed as follows:

$$\dot{N}(M_{\rm BH}) = \Phi(M_{\rm BH}) \cdot \Gamma(M_{\rm BH}). \quad (24)$$

Here $\Gamma(M_{\rm BH})$ is the TDE rate per galaxy (in units of yr$^{-1}$ galaxy$^{-1}$) as a function of $M_{\rm BH}$, and $\Phi(M_{\rm BH})$ is the BHMF, which gives the number density of BHs as a function of their masses.

We adopt two BHMFs for this calculation. One BHMF is from E. Gallo & A. Sesana (2019), for which T. H. T. Wong et al. (2022) provide a more precise formula:

$$\log_{10}\left(\frac{\Phi_{\rm BH}}{{\rm Mpc}^{-3}\ {\rm dex}^{-1}}\right) = -9.82 - 1.1 \times \log_{10}\left(\frac{M_{\rm BH}}{10^7\ M_\odot}\right)$$
$$- \left(\frac{M_{\rm BH}}{128 \times 10^7\ M_\odot}\right)^{1/\ln(10)}. \quad (25)$$

The other BHMF used is from H. Cho & J.-H. Woo (2024), and more specifically we use their bpl-mS model, which is described as follows:

$$\log_{10}\left(\frac{\Phi_{\rm BH}}{{\rm Mpc}^{-3}{\rm dex}^{-1}}\right) = \log_{10}(\phi_\bullet \cdot p(M_{\rm BH}|\theta) \cdot M_{\rm BH}) \quad (26)$$

$$p(M_{\rm BH}|\theta) \sim \frac{1}{M_{\rm BH}}\left(\frac{M_{\rm BH}}{M_c}\right)^\alpha \left(\frac{1}{2}[1+\frac{M_{\rm BH}}{M_c}]\right)^{\beta-\alpha}, \quad (27)$$

with $\phi_\bullet = 2.45 \times 10^{-4}\ {\rm Mpc}^{-3}$, $M_c = 10^{6.75} M_\odot$, $\alpha = -0.60$, $\beta = -2.56$. This BHMF is computed using an observed type 1 AGN sample in both SMBH and IMBH regimes. As TDEs can be produced by both AGNs and quiescent MBHs, under the assumption that AGNs and quiescent MBHs have the same mass distributions, we obtain the BMHF for the MBH population as a whole by first dividing their BHMF by the type 1 fraction of AGNs (fraction of type 1 AGNs among all AGNs) and then by the AGN duty cycle (fraction of active MBHs among all MBHs). We adopt the average duty cycle of 13% and overall type 1 fraction of 40% from H. Cho & J.-H. Woo (2024), but we note that the type 1 fraction is not well constrained in the IMBH regime, which might get as low as 7%. Hence, it is possible that we have underestimated the BHMF in the IMBH regime by using an average type 1 fraction.

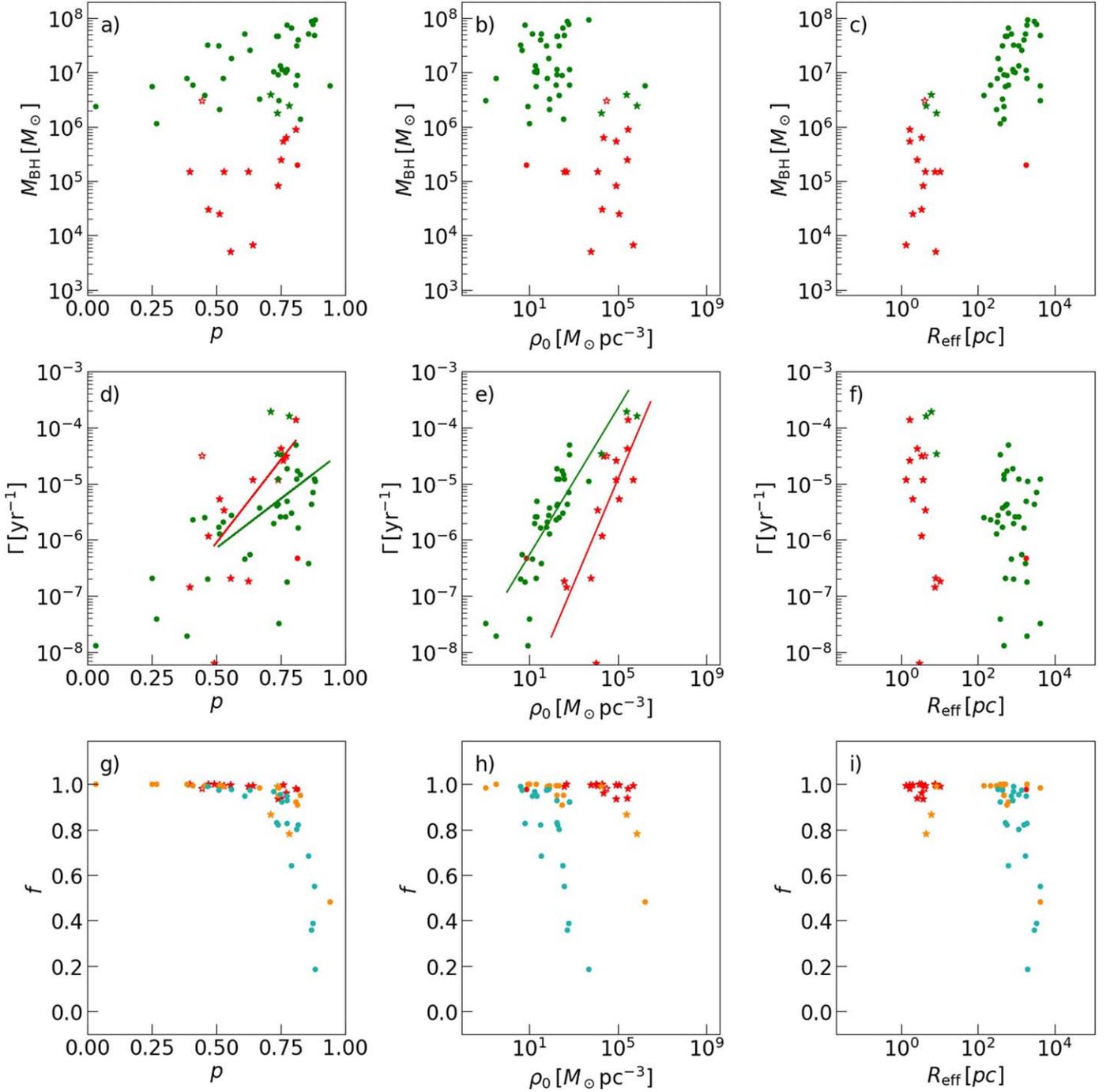

**Figure 4.** MBH mass $M_{\rm BH}$, individual galaxy TDE rate $\Gamma$, and TDE pinhole fraction $f$ (from top to bottom) plotted against the Sérsic inner slope $p$, central density $\rho_0$, and effective radius $R_{\rm eff}$ (from left to right) for our IMBH and SMBH samples. The symbol style and colors are consistent with Figure 1(d), except in panels (g), (h), and (i), where the SMBHs are further divided into two subgroups: $M_{\rm BH} > 10^7\,M_\odot$ (turquoise) and $10^6\,M_\odot < M_{\rm BH} < 10^7\,M_\odot$ (orange). In panels (d) and (e), the red and green lines also display the fitting functions (Equations (22) and (23)) for IMBH-GN hosts and SMBH hosts, respectively.

We plot these two BHMFs by E. Gallo & A. Sesana (2019) and H. Cho & J.-H. Woo (2024) in Figure 5(a). One can see that the BHMF by E. Gallo & A. Sesana (2019) monotonically decreases with $M_{\rm BH}$, almost like a power-law function, while the BHMF by H. Cho & J.-H. Woo (2024) peaks around $M_{\rm BH} = 10^6\,M_\odot$ and then decreases toward lighter or heavier BHs.

We further note that both BHMFs are derived from MBHs observed from galaxy nuclei and hence do not account for IMBHs in GCs. We therefore can only compute the volumetric rates of TDEs from galaxy nuclei based on them. To be specific, we take the binned average TDE rate $\Gamma(M_{\rm BH})$ for the IMBH-GN and SMBH populations (similar to the histogram shown in Figure 1(b) but taking out the contribution from IMBH-ONs) to calculate $\dot{N}$ based on Equation (24).

The results on both BHMFs are shown in Figure 5(b). One can see that, despite the different intrinsic shapes of the two BMHFs, the two volumetric TDE rate curves have overall similar behaviors and both peak around $M_{\rm BH} = 10^6\,M_\odot$. However, the BHMF by H. Cho & J.-H. Woo (2024) gives slightly higher TDE rates at the peak and slightly lower TDE rates when $M_{\rm BH}$ approaches $10^8 M_\odot$ or $10^6 M_\odot$. Here we fit the volumetric TDE rate as a function of $M_{\rm BH}$ based on the BHMF by E. Gallo & A. Sesana (2019) using a broken power-law function, with the break point set at $M_{\rm BH} = 10^6 M_\odot$:

$$\dot{N} = \begin{cases} 1.7 \times 10^{-7} \left(\dfrac{M_{\rm BH}}{10^6 M_\odot}\right)^{0.6} & {\rm yr}^{-1}\,{\rm Mpc}^{-3}\,{\rm dex}^{-1} \\ & {\rm when}\, M_{\rm BH} \leqslant 10^6 M_\odot \\ 1.7 \times 10^{-7} \left(\dfrac{M_{\rm BH}}{10^6 M_\odot}\right)^{-1.0} & {\rm yr}^{-1}\,{\rm Mpc}^{-3}\,{\rm dex}^{-1} \\ & {\rm when}\, M_{\rm BH} > 10^6 M_\odot \end{cases}.$$
(28)

The total volumetric TDE rates are then estimated to be around $\dot{N}_{\rm IMBH-GN} \approx 1 \times 10^{-7}\,{\rm Mpc}^{-3}\,{\rm yr}^{-1}$ and $\dot{N}_{\rm SMBH} \approx 7 \times 10^{-8}\,{\rm Mpc}^{-3}\,{\rm yr}^{-1}$, which are on the same order of magnitude.

We also obtain a first-order estimation of the volumetric rate of IMBH-ON TDEs as follows. We start from the volumetric rate of the SMBH TDEs $\dot{N}_{\rm SMBH}$ as estimated above, and then we scale it first with the ratio between the average TDE rate from an individual SMBH ($\Gamma_{\rm avg,SMBH} = 1.3 \times 10^{-4}\,{\rm galaxy}^{-1}\,{\rm yr}^{-1}$) and that from an individual IMBH-ON ($\Gamma_{\rm avg,IMBH-ON} = 6.9 \times 10^{-6}\,{\rm galaxy}^{-1}\,{\rm yr}^{-1}$). On average, one massive galaxy can have around $N_{\rm GC} = 10$ GCs (Y. Chen & O. Y. Gnedin 2023). However, only a fraction of GCs should host IMBHs. This occupation fraction $f_{\rm occ}$ is poorly constrained, and we take a range of $f_{\rm occ}$ from 10% (V. L. Tang et al. 2024) to the upper limit of 100%. The volumetric rate of TDEs from IMBH-ONs is therefore estimated to be

$$\dot{N}_{\rm IMBH-ON} = \dot{N}_{\rm SMBH} \times \frac{\Gamma_{\rm avg,IMBH-ON}}{\Gamma_{\rm avg,SMBH}} \times N_{\rm GC} \times f_{\rm occ}$$
$$= 4 \times (10^{-9} - 10^{-8})\ {\rm Mpc}^{-3}\ {\rm yr}^{-1}. \quad (29)$$

This means that the volumetric rate of off-center TDEs from IMBH-ONs can be of the same order of magnitude or one order of magnitude lower compared to TDEs from SMBHs.

One can also compare these theoretical TDE rates with the observed rates, although we note that there is a large discrepancy in the latter as reported in the literature. For example, S. van Velzen (2014) calculated the TDE rates based on surveys from the Sloan Digital Sky Survey and Pan-STARRS and obtained a volumetric rate of $10^{-8}\,{\rm Mpc}^{-3}\,{\rm yr}^{-1}$ for TDEs around SMBHs. In later work, S. Velzen (2018) combined more surveys, including GALEX, PTF, iPTF, and ASAS-SN, and updated this number upward to be $5 \times 10^{-7}\,{\rm Mpc}^{-3}\,{\rm yr}^{-1}$. On the forefront of IMBH TDE observations, D. Lin et al. (2018) reported an off-center TDE candidate J2150-0551 from an IMBH with mass between $5 \times 10^4 M_\odot$ and $10^5 M_\odot$ found by searching the 3XMM-DR5 catalog, and from this they estimated the IMBH TDE volumetric rate to be $\sim 10^{-8}\,{\rm Mpc}^{-3}\,{\rm yr}^{-1}$. Moreover, the Young Supernova Experiment released its first set of data (YSE DR1), which includes five TDEs (P. D. Aleo et al. 2023), of which AT 2020neh is of particular interest, as it was found to be in the vicinity of an IMBH with $M_{\rm BH}$ ranging between $10^{4.7}$ and $10^{5.9} M_\odot$ (C. R. Angus et al. 2022). Based on this, the estimated IMBH TDE rate is $\lesssim 2 \times 10^{-8}\,{\rm Mpc}^{-3}\,{\rm yr}^{-1}$. Very recently, Y. Yao et al. (2023) also reported a TDE rate of $\sim 3 \times 10^{-7}\,{\rm Mpc}^{-3}\,{\rm yr}^{-1}$ using 33 optically selected TDEs from the Zwicky Transient Facility, with $M_{\rm BH}$ ranging from $10^{5.1}$ to $10^{8.2} M_\odot$. TDE rate determinations from larger samples are needed to settle the issue.

We have compiled the more recent observed TDE rate estimates mentioned above in Figure 5. While our theoretical rates and the observed rates of TDEs seem roughly in good agreement, we caution that there are uncertainties in both the observed and theoretical TDE rates. Factors such as selection effect, instrumentation bias, and obscuration should systematically lower the chance of observing a TDE. Furthermore, current transient surveys might not catch all IMBH TDEs, which reside off-center in massive galaxies or have nontypical flaring evolution timescales (J.-H. Chen & R.-F. Shen 2018; H. Pfister et al. 2022). The upshot is that the true TDE rate is likely somewhat higher than the observed rate, and the values reported should be interpreted as a lower limit instead. On the theoretical side, multiple factors also contribute to the uncertainties in the TDE rate calculation. One that is widely debated is the occupation fraction of IMBHs in galaxies, which is expected to vary as a function of epoch and host galaxy mass. The BHMF we used in this work is based on E. Gallo & A. Sesana (2019), which has a $2\sigma$ uncertainty of $\sim \pm 1$ dex for IMBHs in the centers of low-mass galaxies. Moreover, the IMBH mass estimates also bear huge uncertainties, with the differences between the best and maximum $M_{\rm BH}$ estimates reaching as high as 3 dex. Last but not least, TDE physics and stellar dynamics play an important role. For example, when including the effects of mass segregation, it was shown that the TDE rate generally declines over time (L. Broggi et al. 2022). In addition, TDEs in certain parameters might not produce bright or prompt flares, which lowers the chance of their detection given our current observational strategies (C. S. Kochanek 2016; T. H. T. Wong et al. 2022).

### 3.5. The Distribution of the Penetration Parameter $\beta$

In this subsection, we calculate the distribution of the penetration parameter $\beta$ for both IMBH and SMBH TDEs. The probability density function of $\beta$ is calculated as follows:

$$\frac{dP}{d\beta} = \frac{1}{\beta \Gamma} \int \frac{d^2\Gamma}{dE d\ln\beta} dE. \quad (30)$$

We show the results in Figure 6 for our IMBH and SMBH samples. The curves are color-coded by the values of $M_{\rm BH}$, going from green ($10^7$–$10^8 M_\odot$) to red ($10^4$–$10^5 M_\odot$).

The IMBH TDEs and SMBH TDEs with $M_{\rm BH} \lesssim 10^7 M_\odot$ have $\beta$-distributions that largely follow $dP/d\beta \propto \beta^{-2}$, which is consistent with the pinhole-dominated regime. The curves start to drop at very large $\beta$, where the smallest/densest stars need to get within the BH event horizon before they are disrupted, and eventually cut off at a $\beta$ that corresponds to when the biggest/least dense stars also plunge into the BH before being disrupted. On the other hand, when $M_{\rm BH} \gtrsim 10^7 M_\odot$, the $\beta$-distribution is steeper, indicating that some fraction of TDEs can also be produced in the diffusive regime.

Interestingly, TDEs from IMBHs can have $\beta$ as high as $\sim 100$–1000. This is very different from what is expected for SMBH TDEs, for which $\beta$ can only reach around 10. Observationally, it has been reported that very high $\beta$ might cause the flare rise time to become longer (J. Guillochon & E. Ramirez-Ruiz 2013). In an extreme case, it has been argued that the compression experienced by a very deeply plunging star might even be strong enough to trigger nuclear burning, and the explosion induced can resemble a Type Ia supernova and produce unique signatures (S. Rosswog et al. 2009).

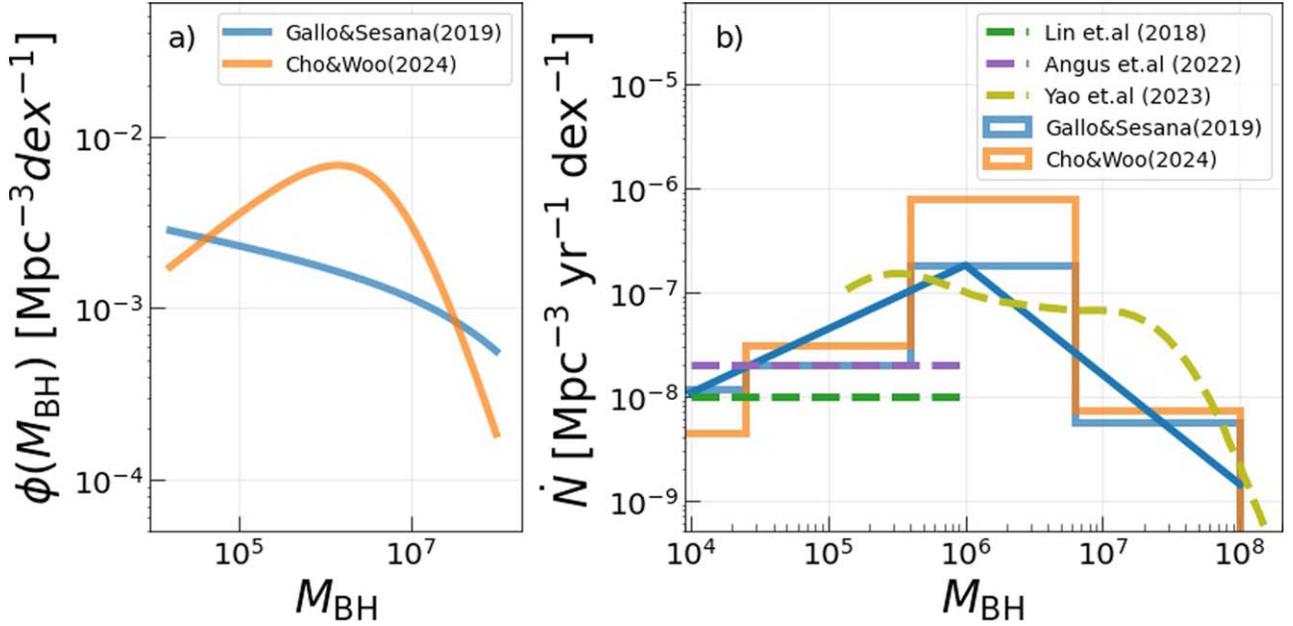

**Figure 5.** The BHMFs and the volumetric TDE rate. Panel (a) shows the BHMF from E. Gallo & A. Sesana (2019) in blue and that from H. Cho & J.-H. Woo (2024) (after the correction with the AGN duty cycle and type 1 fraction) in orange. Panel (b) shows the volumetric TDE rate $\dot{N}$ vs. $M_{BH}$. The histograms are obtained from convolving the binned average of the IMBH-GN and SMBH TDE rates (similar to the histogram shown in Figure 1(b)) with the BHMFs shown in panel (a). The blue line shows the fitting function (Equation (28)) for the volumetric rate computed using the BHMF from E. Gallo & A. Sesana (2019). Comparison to observations is also shown using dashed lines: D. Lin et al. (2018) and C. R. Angus et al. (2022) reported volumetric IMBH TDE rates of $\sim 10^{-8}$ Mpc$^{-3}$ yr$^{-1}$ based on the observation of J2150-055 and AT 2020neh, respectively, and Y. Yao et al. (2023) reported the volumetric TDE rate as a function of $M_{BH}$ based on the 33 optically selected ZTF TDEs.

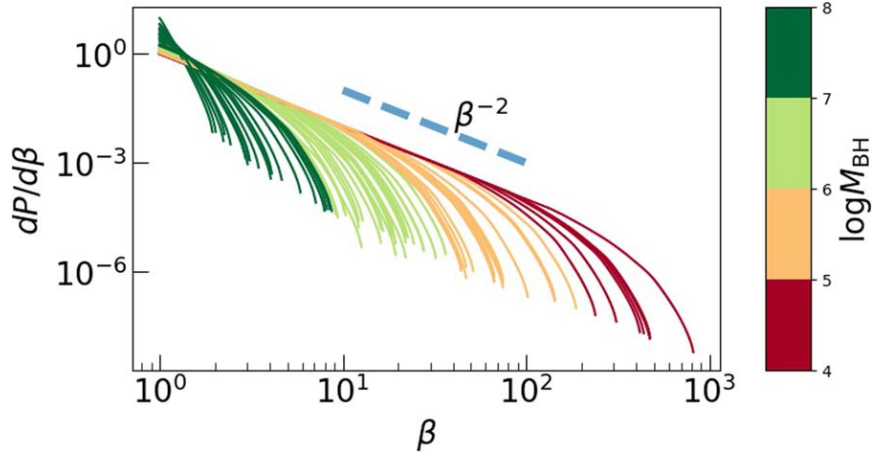

**Figure 6.** The distribution of penetration parameter $\beta$ in TDEs for different $M_{BH}$. The curves are color-coded by $M_{BH}$, from $10^8\,M_\odot$ (green) to $10^4\,M_\odot$ (red). The dashed line denotes the $dP/d\beta \propto \beta^{-2}$ relation that characterizes the pinhole-dominated regime.

### 3.6. A Detailed Investigation on How BH Mass and Galaxy Stellar Profile Affect the TDE Rate

We delve deeper to investigate the behavior of the TDE rate $\Gamma$ computed from individual galaxies as a function of $M_{BH}$ described in Section 3.1 and seek to understand the cause for the $\Gamma$ versus $M_{BH}$ trend as shown in Figure 1(b). To explore this, we first compute the TDE rate for each galaxy or cluster by varying $M_{BH}$ as a free parameter over a range of $10^3$–$10^8\,M_\odot$ while keeping the galaxy stellar profiles fixed as dictated by observations. The $\Gamma$ versus $M_{BH}$ results for IMBHs and SMBHs are presented in Figure 7, where the curves show the fit using a third-degree polynomial. Interestingly, for almost all galaxies in our sample, $\Gamma$ from each galaxy exhibits a similar behavior, in that it first increases with $M_{BH}$ and then turns over at a certain $M_{BH}$, after which $\Gamma$ reverses and starts to decrease with $M_{BH}$.

We further examine the value of the optimal BH mass ($M_{BH,peak}$) at which the $\Gamma(M_{BH})$ curve reaches maximum and compare that to the actual observed BH mass ($M_{BH,obs}$) for each galaxy. Figure 8 shows the difference between these two masses ($\Delta \log M_{BH} \equiv \log M_{BH,peak} - \log M_{BH,obs}$) as a function of $\log M_{BH,obs}$. We see a monotonically decreasing trend between $\Delta \log M_{BH}$ and $\log M_{BH,obs}$, and the sign of $\Delta \log M_{BH}$ switches at around a few times $10^6\,M_\odot$. This, again, demonstrates that $\Gamma(M_{BH})$ should itself

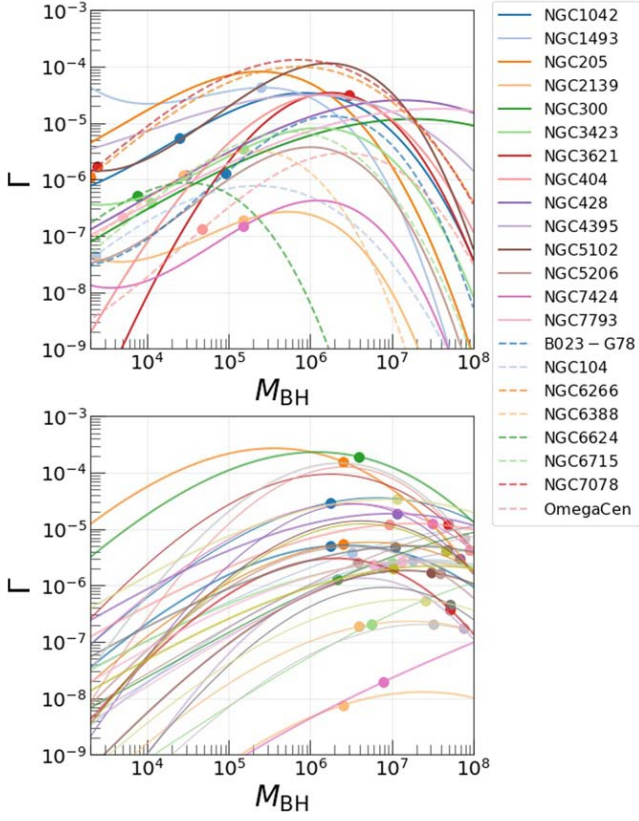

Figure 7. TDE rate $\Gamma$ from each galaxy or cluster as a function of $M_{BH}$ (while keeping the stellar profiles fixed) from IMBHs (top panel) and SMBHs (bottom panel). Solid lines and dotted lines in the top panel represent IMBH-GNs and IMBH-ONs, respectively. The circles mark the observed BH masses $M_{BH,obs}$ and the corresponding TDE rates.

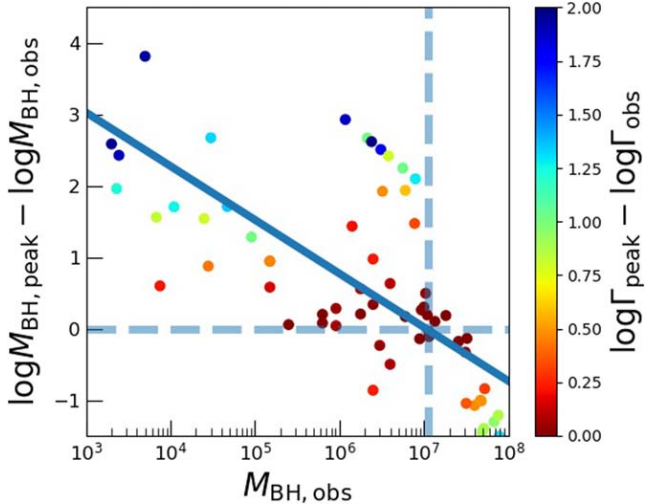

Figure 8. The difference between the observed BH mass $M_{BH,obs}$ and the BH mass at which $\Gamma(M_{BH})$ reaches maximum $M_{BH,peak}$, plotted against $M_{BH,obs}$, for each galaxy or cluster in our sample. The color indicates the difference in TDE rates calculated at $M_{BH,peak}$ and $M_{BH,obs}$. A linear fit is shown using the solid blue line. The dashed blue lines indicate where the y-axis is 0 (i.e., where $\Delta \log M_{BH}$ switches sign), which crosses the linear fit at $\log M_{BH,obs} \sim 7$.

on average also peak at around $10^6 M_\odot$ for the galaxies/clusters in our sample.

Next, we check how the galaxy stellar profile affects the computation of $\Gamma(M_{BH})$. We plot $d\log\Gamma/d\log M_{BH}$ against the Sérsic inner slope $p$ for each galaxy in Figure 9. Here $d\log\Gamma/d\log M_{BH}$ is the slope of $\Gamma(M_{BH})$ calculated at the observed BH mass $M_{BH,obs}$. On average, we find that $d\log\Gamma/d\log M_{BH}$ decreases with $p$ (although some scatter exists), which is consistent with the results in J. Wang & D. Merritt (2004). Around a value of $p \sim 0.6$, the sign of $d\log\Gamma/d\log M_{BH}$ switches from positive to negative. This means that when the stellar component is less centrally concentrated (lower value of $p$), $\Gamma(M_{BH})$ tends to increase with $M_{BH}$, and most IMBH-GN hosts reside within this regime. On the other hand, for a more centrally concentrated stellar component (higher value of $p$), $\Gamma(M_{BH})$ more likely decreases with $M_{BH}$. Our SMBH host population roughly splits between these two $p$-regimes.

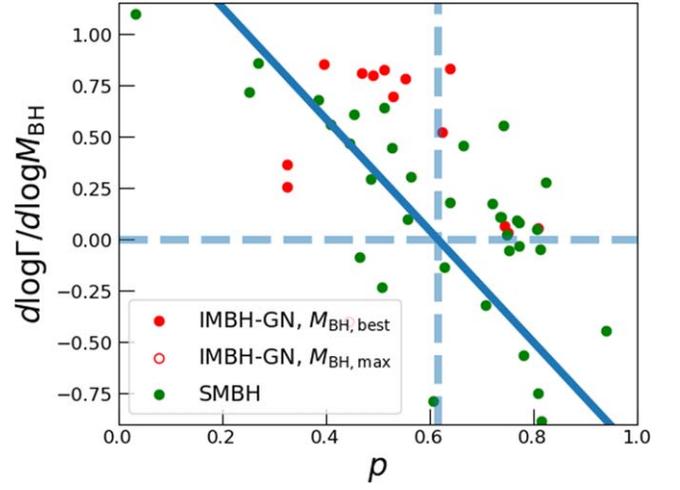

Figure 9. The gradient of $\Gamma(M_{BH})$ calculated at the observed $M_{BH,obs}$ plotted against the Sérsic inner slope $p$ (of the innermost stellar component) for each galaxy. The circle symbol style and color are consistent with Figure 1(a). The solid blue line shows the linear fit, and the dashed blue line marks where $d\log\Gamma/d\log M_{BH} = 0$. The curves cross at $p \sim 0.62$.

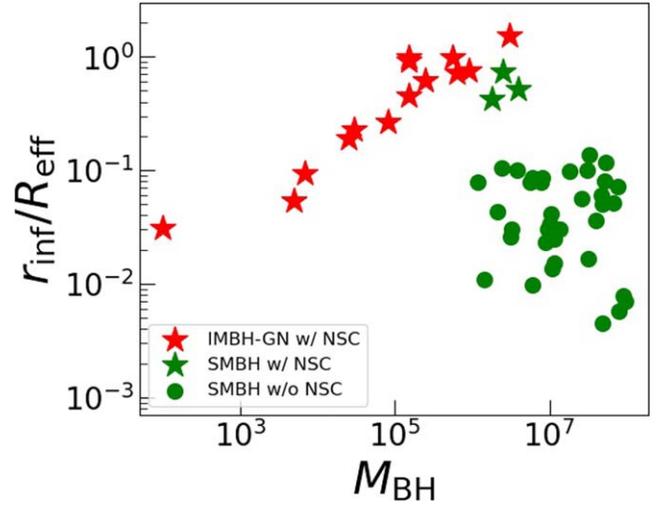

Figure 10. The ratio between the BH influence radius $r_{inf}$ and the effective radius $R_{eff}$ (of the innermost stellar component) of a galaxy plotted against $M_{BH}$ for each galaxy. The symbol style and color are consistent with Figure 1(d).

Finally, we investigate how the effective size $R_{eff}$ of the innermost stellar component of a galaxy affects $\Gamma$. To be more specific, in Figure 10 we plot the ratio between the BH influence radius $r_{inf}$ and the effective radius of the inner stellar component $R_{eff}$ against $M_{BH}$. Interestingly, the trend of this ratio also somewhat resembles the trend seen in $\Gamma$ versus $M_{BH}$ (Figure 1(b)): $r_{\rm inf}/R_{\rm eff}$ initially increases with $M_{BH}$ at low $M_{BH}$ and then decreases with $M_{BH}$ at high $M_{BH}$ with a turnover at $\sim 10^6 \, M_\odot$. We take a closer look at this behavior and offer the following explanation: (1) In the IMBH mass range, as $M_{BH}$ increases, $r_{\rm inf}$ also increases, while $R_{\rm eff}$ of the NSCs does not vary much (as shown in Figure 4 panel (c)); a larger $r_{\rm inf}$ therefore puts more stars within the gravitational influence of the BH, which enhances the TDE rate. (2) Parameter $r_{\rm inf}$ becomes comparable to $R_{\rm eff}$ around $M_{BH} = 10^6 \, M_\odot$, the mass scale that separates our IMBH and SMBH populations. Beyond this $M_{BH}$, $r_{\rm inf}$ still increases with $M_{BH}$. However, the $R_{\rm eff}$ of SMBH galaxy bulges are much larger than the sizes of the compact NSCs, so $r_{\rm inf}/R_{\rm eff}$ overall drops to lower values. A lower $r_{\rm inf}/R_{\rm eff}$ means that the stars are more spread out (compared to the sphere of influence of the BH), which then decreases the TDE rate.

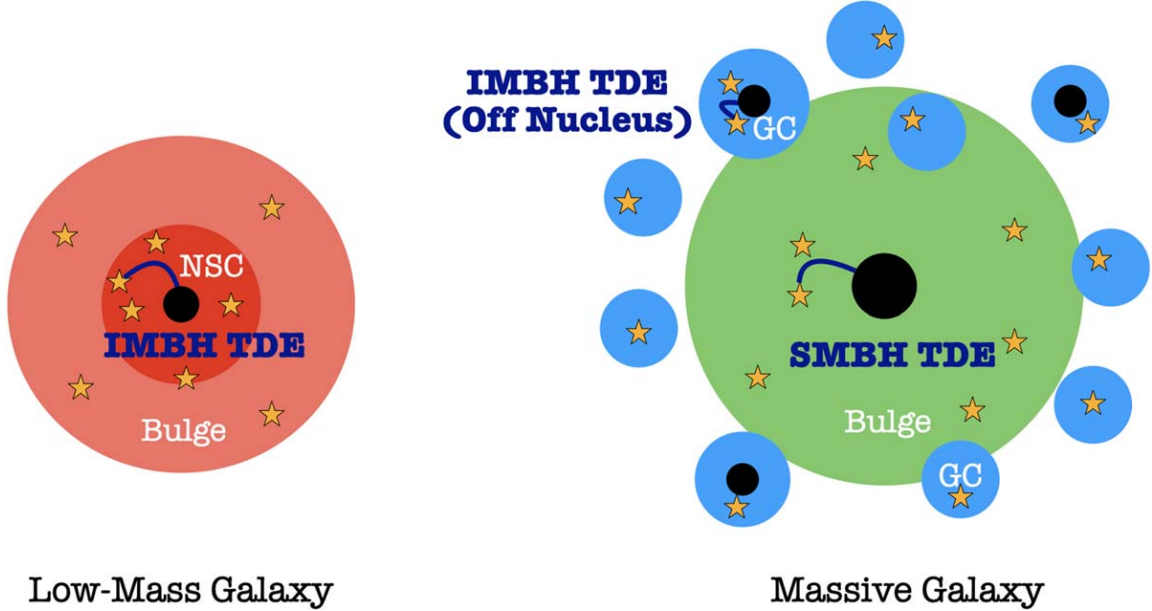

**Figure 11.** A schematic of two types of galaxies and the TDEs occurring in these galaxies. Left: a low-mass galaxy with an IMBH residing at the galactic nucleus. Such a galaxy typically has not only a stellar bulge but also denser NSC components, which boost TDE rates. A TDE can occur around the central IMBH (i.e., TDE IMBH-GN). Right: a massive galaxy with an SMBH at the galactic nucleus. The massive galaxy is surrounded by GCs, some of which can host IMBHs. TDEs can occur around the SMBH at the center of the main galaxy (i.e., TDE SMBH), or around the IMBHs that are off nuclei (i.e., TDE IMBH-ON).

In summary, we conclude that the opposite trend of $\Gamma$ versus $M_{BH}$ seen in the IMBH population to that of the SMBH population is a result of a combination of factors. For most realistically observed stellar profiles, $\Gamma(M_{BH})$ peaks around the sweet spot of $M_{BH} \sim 10^6$–$10^7 \, M_\odot$. In addition, the stellar components of the IMBH and SMBH host galaxies are structured differently in terms of both $p$ and $R_{\rm eff}$, which contributes to having $\Gamma(M_{BH})$ peak around the mass separation between the two populations.

## 4. Conclusion

In this work we compute the TDE rates and pinhole fractions for a sample of 22 observed IMBH candidates and 40 nearby SMBHs, which all have realistic stellar profiles for their host galaxies and BH masses constrained using dynamical measurements. Some of these IMBHs reside at the center of small galaxies (denoted as IMBH-GN), while others reside in GCs (denoted as IMBH-ON). We illustrate these three different populations of TDEs and their hosts using a schematic in Figure 11.

We compare the results of our TDE calculations for IMBHs and SMBHs and investigate how the TDE rate and pinhole fraction are affected by the host galaxy/cluster stellar profile and the BH mass. Furthermore, we also calculate the $\beta$-distributions and the volumetric rates of IMBH and SMBH TDEs. We summarize the key results below and present some additional technical details and derivations in the appendices:

1. We find that the TDE rate for an IMBH-hosting galaxy or cluster in our sample varies from $\sim 10^{-8}$ to $\sim 10^{-4} \, {\rm galaxy}^{-1} \, {\rm yr}^{-1}$, which is comparable to the TDE rate for an SMBH-hosting galaxy (Figure 1). This implies that TDEs can serve as a promising probe for the detection of IMBHs, which remain a somewhat elusive population despite expectations of them being ubiquitous. Among the two IMBH populations, the TDE rate for an IMBH-GN is typically higher than that for an IMBH-ON.
2. Contrary to the predictions from earlier works, we find that the TDE rate does not monotonically decrease when $M_{BH}$ increases. The TDE rate from every galaxy generally increases as $M_{BH}$ increases within the IMBH mass range of $10^4$–$10^6 \, M_\odot$, while it decreases as $M_{BH}$ moves into the SMBH regime for BH mass in excess of $10^6 \, M_\odot$ (as shown in Figure 1(b)). We have provided a fitting formula for the TDE rate per galaxy as a function of $M_{BH}$ (Equation (18)).
3. IMBH TDEs mostly occur in the pinhole regime (Figure 2). This implies that IMBH TDEs, as compared to SMBH TDEs, can more likely have deeply penetrating events (Figure 6) and produce distinct observational signatures.
4. The TDE rate for a galaxy is positively correlated to both its central stellar density $\rho_0$ and the inner slope of the stellar density profile $p$ (Figure 4) when modeling the

   stellar component of the host galaxy with the Sérsic model. Hence, we see that IMBHs embedded in a galaxy with an NSC have a boosted TDE rate.

5. The TDE pinhole fraction for a galaxy, on the other hand, depends mainly on the inner slope of the stellar profile $p$ (Figure [4](#)). IMBH hosts tend to have low $p$ (i.e., they have less concentrated stellar structures), which explains once again why IMBH TDEs mostly occur in the pinhole regime.

6. We compute the volumetric TDE rates as a function of $M_{BH}$ around IMBHs and SMBHs (Equation ([28](#))) and show that the total volumetric TDE rates around IMBHs and SMBHs in galaxy nuclei are comparable, with $\dot{N}_{IMBH-GN} \approx 1 \times 10^{-7}$ Mpc$^{-3}$ yr$^{-1}$ and $\dot{N}_{SMBH} \approx 7 \times 10^{-8}$ Mpc$^{-3}$ yr$^{-1}$, respectively. We also provide a first-order estimate of the volumetric TDE rates of off-center IMBHs $\dot{N}_{IMBH-ON} \approx 4 \times (10^{-9}$ to $10^{-8})$ Mpc$^{-3}$ yr$^{-1}$, assuming that 10%–100% of GCs host IMBHs. Therefore, the total rate of off-center IMBH TDEs is comparable to TDEs from galactic nuclei. In general, these theoretically estimated TDE rates are in agreement with the estimates derived from current observations.

We also note that the volumetric TDE rate obtained in this work depends sensitively on the assumed BHMF, which has a large degree of uncertainty in the IMBH mass range. However, based on two different BHMFs that are obtained using either AGNs or a mix of AGNs and quiescent MBHs, we find that the volumetric rate of TDEs should always peak around $10^6 M_\odot$. In the future when a larger sample of IMBH TDEs is obtained, it will become possible to use the observed IMBH TDE rates to constrain the IMBH mass function, occupation fraction, and AGN duty cycle conversely. A similar approach has been demonstrated by Y. Yao et al. ([2023](#)), but mainly for SMBHs.

Last but not least, it is intriguing to ask why the number of detected IMBH TDEs is lower than SMBH TDEs, if their intrinsic rates are in fact comparable. Up to now only about a dozen observed TDEs are from MBHs with $M_{BH} < 10^6 M_\odot$ (T. Wevers et al. [2017](#), [2019](#)). One possible reason for the low detection rate could be that quiescent MBHs have a mass distribution very different from AGNs and that the actual IMBH occupation fraction in this BH population is quite low. Alternatively, it is also possible that IMBH TDEs may have characteristics different from classical predictions and that current surveys tend to miss them. Moreover, IMBH TDEs should come from small galaxies and stellar clusters that are off the center of massive, bright galaxies. This adds difficulty in constraining the host properties and identifying the nature of these events. Besides, many automated pipelines for transient detection, including TDEs, currently filter out off-center events, therefore preferentially missing IMBH TDEs. In fact, TDEs might serve to be an efficient way to detect off-center dormant IMBHs or the population that is accreting at very low rates (A. Ricarte et al. [2021a](#), [2021b](#)). However, such sources are also not expected to have an associated dense NSC (P. Natarajan et al. 2024, in preparation). In general, a better understanding of the physics and environments of IMBH TDEs will be very useful for detecting this population of transients more effectively in the future and leveraging them to probe the elusive IMBH population.

## Acknowledgments

We thank H. Cho, D. French, J. Greene, C. Hannah, S. Li, M. Polkas, E. Ramirez-Ruiz, A. Seth, N. Stone, V. Tang, T. H. Wong, and J. Woo for useful discussions. We also thank the anonymous referee for constructive comments. J.C., L.D., H.F., and R.K.C. acknowledge the support from the National Natural Science Foundation of China (HKU12122309) and the Hong Kong Research Grants Council (HKU17314822, 27305119, 17304821). This research was supported in part by grant No. NSF PHY-2309135 to the Kavli Institute for Theoretical Physics (KITP). P.N. acknowledges support from the Gordon and Betty Moore Foundation and the John Templeton Foundation, which fund the Black Hole Initiative (BHI) at Harvard University, where she serves as one of the PIs.

## Appendix A
## Samples

We present here our sample of IMBH-GNs (Table [A1](#)), IMBH-ONs (Table [A2](#)), and SMBHs (Table [A3](#)).

We also report the uncertainties in the SMBH and "best" IMBH mass measurements and the resultant TDE rate in these tables. For IMBHs, we take the $1\sigma$ values for the $M_{BH,best}$ measurements from D. D. Nguyen et al. ([2019](#)) if available. For the IMBHs from N. Neumayer & C. J. Walcher ([2012](#)), since they did not report the error bars of the BH mass measurements, we adopt a one-sided error bar using $M_{BH,max}$ as the upper bound. For the special case of NGC 4395, since we take the average of the BH mass measurements from the dynamical and reverberation mapping methods, we use the individual mass measurement from these two methods as the lower and upper bounds. For SMBH mass measurements, some already have error bars reported by D. D. Nguyen et al. ([2018](#)) and B. L. Davis et al. ([2019](#)). For the other SMBHs, we estimate the uncertainty in their BH mass measurements using the scattering in the $M$–$\sigma$ relation (N. J. McConnell & C.-P. Ma [2013](#)) based on the $\sigma$ reported by T. R. Lauer et al. ([2007](#)).

Table A1
IMBHs at Galaxies Nuclei

| Name | $n$ | $R_{\rm eff}$ (pc) | $\log_{10}\left(\frac{\rho_0}{M_\odot\,{\rm pc}^{-3}}\right)$ | $\log_{10}\left(\frac{M_{\rm BH,best}}{M_\odot}\right)$ | $\log_{10}\left(\frac{M_{\rm BH,max}}{M_\odot}\right)$ | $\log_{10}\left(\frac{M_\star}{M_\odot}\right)$ | $f$ | $\log_{10}\Gamma$ | Reference |
|---|---|---|---|---|---|---|---|---|---|
| NGC 5102$_1$ | 0.80 | 1.60 | 5.44 | 5.96 | $0.02^{+0.03}_{-6.85}$ | | 1.00 | $-3.99^{+0.01}_{-0.01}$ | 1 |
| NGC 5102$_2$ | 3.10 | 32.00 | 4.19 | 5.96 | $0.02^{+0.03}_{-7.76}$ | | 0.93 | $-4.45^{+0.07}_{0.00}$ | 1 |
| NGC 5102$_b$ | 3.00 | 1200.00 | 1.39 | 5.96 | $0.02^{+0.03}_{-9.77}$ | | 0.97 | $-5.66^{+0.01}_{-0.01}$ | 1 |
| NGC 1493 | 2.36 | 2.60 | 5.41 | 5.40 | 5.90 | 6.30 | 0.94 | $-4.37_{-0.12}$ | 2 |
| NGC 3621 | 1.00 | 4.10 | 4.50 | | 6.48 | 7.00 | 0.98 | $-4.51$ | 3 |
| NGC 5206$_1$ | 0.80 | 3.40 | 3.83 | 5.80 | $0.05^{+0.13}_{-6.23}$ | | 1.00 | $-5.42^{+0.00}_{-0.09}$ | 1 |
| NGC 5206$_2$ | 2.30 | 10.50 | 4.35 | 5.80 | $0.05^{+0.13}_{-7.11}$ | | 0.95 | $-4.57^{+0.02}_{-0.00}$ | 1 |
| NGC 5206$_b$ | 2.57 | 986.00 | 0.92 | 5.80 | $0.05^{+0.13}_{-9.38}$ | | 0.98 | $-6.45^{+0.03}_{-0.08}$ | 1 |
| NGC 404$_1$ | 0.50 | 1.60 | 4.93 | 5.74 | $0.24^{+0.51}_{-6.53}$ | | 1.00 | $-4.63^{+0.10}_{-0.44}$ | 4 |
| NGC 404$_2$ | 1.96 | 20.10 | 3.17 | 5.74 | $0.24^{+0.51}_{-7.04}$ | | 0.98 | $-5.63^{+0.03}_{-0.17}$ | 4 |
| NGC 404$_b$ | 2.43 | 640.00 | 0.92 | 5.74 | $0.24^{+0.51}_{-8.93}$ | | 0.99 | $-6.58^{+0.10}_{-0.25}$ | 4 |
| NGC 4395 | 2.25 | 3.60 | 4.90 | 4.92 | $0.68^{+0.68}_{-6.30}$ | | 0.94 | $-4.92^{+0.39}_{-0.10}$ | 5,6 |
| NGC 4395$_b$ | 0.44 | 500.00 | 0.18 | 4.92 | $0.68^{+0.68}_{-9.30}$ | | 1.00 | $-9.50^{+1.25}_{-3.46}$ | 5,6 |
| NGC 205 | 1.60 | 1.30 | 5.68 | 3.83 | $0.43^{+0.60}_{-6.26}$ | | 1.00 | $-4.92^{+0.35}_{-0.48}$ | 1 |
| NGC 205$_b$ | 1.40 | 516.00 | 0.47 | 3.83 | $0.00^{+0.00}_{-8.99}$ | | 1.00 | $-8.02^{+0.05}_{-0.59}$ | 1 |
| NGC 1042 | 1.15 | 1.94 | 5.05 | 4.40 | 6.48 | 6.48 | 1.00 | $-5.27^{+0.64}$ | 2 |
| NGC 3423 | 1.20 | 4.18 | 4.09 | 5.18 | 5.85 | 6.48 | 1.00 | $-5.46^{+0.34}$ | 2 |
| NGC 428 | 1.05 | 3.36 | 4.27 | 4.48 | 4.85 | 6.48 | 1.00 | $-5.92^{+0.29}$ | 2 |
| NGC 7793 | 1.27 | 7.70 | 3.77 | 3.70 | 5.90 | 6.90 | 1.00 | $-6.68^{+1.49}$ | 2 |
| NGC 2139 | 1.53 | 10.30 | 2.58 | 5.18 | 5.60 | 5.90 | 0.99 | $-6.73^{+0.15}$ | 2 |
| NGC 7424 | 0.91 | 7.40 | 2.66 | 5.18 | 5.60 | 6.00 | 1.00 | $-6.83^{+0.31}$ | 2 |
| NGC 300 | 1.10 | 2.90 | 4.02 | 2.00 | 5.00 | 6.00 | 1.00 | $-8.20^{+2.42}$ | 2 |

**Note.** Table of stellar profile parameters for IMBH-GNs. The parameters are (from left to right) the Sérsic index $n$, the effective radius $R_{\rm eff}$, central density $\rho_0$, best BH mass estimate $M_{\rm BH,best}$, maximum BH mass estimate $M_{\rm BH,max}$, mass of the stellar component $M_\star$, pinhole fraction $f$, and TDE rate $\Gamma$. The table is sorted by TDE rate from high to low. Some galaxies have NSCs and therefore are fitted with multicomponent stellar profiles. These components are indicated by different lower indices, with numbers indicating the NSC component and b indicating the bulge component. The stellar profiles and BH masses are obtained from the following sources: (1) D. D. Nguyen et al. (2019), (2) N. Neumayer & C. J. Walcher (2012), (3) A. J. Barth et al. (2009), (4) T. A. Davis et al. (2020), (5) M. den Brok et al. (2015), (6) H. Cho et al. (2021), and (7) C. R. Angus et al. (2022). For the special case of NGC 4395, which has mass measurements from two different methods (dynamical measurements giving $\log_{10}(M_{\rm BH}/M_\odot) = 5.60$ (M. den Brok et al. 2015) and reverberation mapping giving $\log_{10}(M_{\rm BH}/M_\odot) = 4.23$ (J.-H. Woo et al. 2019; H. Cho et al. 2021)), we calculate the geometric mean of the two masses and use that to calculate the TDE rate and pinhole fraction.

Table A2
IMBHs Off Nuclei

| Name | $W_0$ | $R_0$ | $\log_{10}\left(\frac{M_{\rm BH,max}}{M_\odot}\right)$ | $\log_{10}\left(\frac{M_\star}{M_\odot}\right)$ | $f$ | $\log_{10}\Gamma$ | Reference |
|---|---|---|---|---|---|---|---|
| NGC 7078 | 10.82 | 0.20 | 3.39 | 6.71 | 1.00 | $-5.77$ | 1 |
| B023-G78 | 5.37 | 2.69 | 4.96 | 6.79 | 1.00 | $-5.90$ | 2 |
| NGC 6388 | 11.00 | 0.41 | 4.45 | 6.08 | 1.00 | $-5.92$ | 3 |
| NGC 6266 | 7.59 | 0.21 | 3.30 | 5.85 | 1.00 | $-5.96$ | 4 |
| NGC 6624 | 10.82 | 0.13 | 3.88 | 4.86 | 1.00 | $-6.29$ | 5 |
| NGC 6715 | 7.98 | 0.81 | 4.04 | 6.15 | 1.00 | $-6.42$ | 6 |
| OmegaCen | 5.50 | 4.60 | 4.67 | 6.70 | 1.00 | $-6.89$ | 7 |
| NGC 104 | 12.00 | 0.42 | 3.36 | 5.89 | 1.00 | $-7.35$ | 8 |

**Note.** Table of stellar profile parameters for IMBH-ONs. The parameters are (from left to right) the dimensionless potential $W_0$, the King radius $R_0$, BH mass estimate $M_{\rm BH}$, stellar mass $M_\star$, pinhole fraction $f$, and TDE rate $\Gamma$. The table is sorted by TDE rate. The stellar profiles and BH mass are obtain from the following sources: (1) J. Gerssen et al. (2002), (2) R. Pechetti et al. (2022), (3) N. Lützgendorf et al. (2015), (4) N. Lützgendorf et al. (2013), (5) B. B. P. Perera et al. (2017), (6) H. Baumgardt (2017), (7) E. Noyola et al. (2010), and (8) B. Kızıltan et al. (2017).

Table A3
SMBHs

| Name | $n$ | $R_{\mathrm{eff}}$ (pc) | $\log_{10}\left(\frac{\rho_0}{M_\odot\,\mathrm{pc}^{-3}}\right)$ | $\log_{10}\left(\frac{M_{\mathrm{BH}}}{M_\odot}\right)$ | $\log_{10}\left(\frac{M_\star}{M_\odot}\right)$ | $f$ | $\log_{10}\Gamma$ |
|---|---|---|---|---|---|---|---|
| Circinus$_1$ | 2.21 | 680.00 | 1.86 | $6.25^{+0.12}_{-0.10}$ | 10.12 | 0.98 | $-5.30^{+0.01}_{-0.01}$ |
| Circinus$_2$ | 1.09 | 8.00 | 4.26 | $6.25^{+0.12}_{-0.10}$ | 7.57 | 0.99 | $-4.53^{+0.03}_{-0.03}$ |
| ESO558−G009 | 1.28 | 330.00 | 1.87 | $7.26^{+0.04}_{-0.03}$ | 9.89 | 0.98 | $-5.56^{+0.00}_{-0.00}$ |
| IC 2560 | 2.27 | 4210.00 | −0.96 | $6.49^{+0.21}_{-0.19}$ | 9.63 | 0.98 | $-7.48^{+0.12}_{-0.10}$ |
| J0437+2456 | 1.73 | 420.00 | 1.90 | $6.51^{+0.05}_{-0.04}$ | 9.90 | 0.98 | $-5.43^{+0.02}_{-0.02}$ |
| M32$_1$ | 2.70 | 4.40 | 5.85 | $6.40^{+0.20}_{-0.10}$ | 7.16 | 0.78 | $-3.80^{+0.05}_{-0.13}$ |
| M32$_2$ | 1.60 | 108.00 | 2.57 | $6.40^{+0.20}_{-0.10}$ | 8.90 | 0.98 | $-5.26^{+0.02}_{-0.02}$ |
| M32$_3$ | 1.00 | 516.00 | −0.52 | $6.40^{+0.20}_{-0.10}$ | 8.29 | 1.00 | $-8.12^{+0.08}_{-0.05}$ |
| Milky Way$_1$ | 1.30 | 1040.00 | 0.46 | $6.60^{+0.02}_{-0.02}$ | 9.96 | 0.99 | $-6.73^{+0.01}_{-0.01}$ |
| Milky Way$_2$ | 2.00 | 6.00 | 5.38 | $6.60^{+0.02}_{-0.02}$ | 7.64 | 0.87 | $-3.72^{+0.01}_{-0.01}$ |
| Mrk 1029 | 1.15 | 300.00 | 1.91 | $6.33^{+0.13}_{-0.10}$ | 9.90 | 1.00 | $-5.89^{+0.08}_{-0.06}$ |
| NGC 0253 | 2.53 | 930.00 | 1.34 | $7.00^{+0.30}_{-0.30}$ | 9.76 | 0.95 | $-5.59^{+0.01}_{-0.04}$ |
| NGC 1068 | 0.71 | 510.00 | 1.29 | $6.75^{+0.08}_{-0.08}$ | 10.27 | 1.00 | $-6.68^{+0.06}_{-0.06}$ |
| NGC 1300 | 4.20 | 1720.00 | 1.55 | $7.71^{+0.14}_{-0.19}$ | 9.42 | 0.69 | $-6.42^{+0.24}_{-0.20}$ |
| NGC 1320 | 3.08 | 610.00 | 2.82 | $6.78^{+0.34}_{-0.24}$ | 10.25 | 0.92 | $-4.30^{+0.00}_{-0.02}$ |
| NGC 1426 | 4.95 | 4150.00 | 2.57 | $7.69^{+0.11}_{-0.06}$ | 10.98 | 0.55 | $-4.91^{+0.08}_{-0.15}$ |
| NGC 2273 | 2.24 | 460.00 | 2.25 | $6.97^{+0.09}_{-0.09}$ | 9.98 | 0.95 | $-4.92^{+0.01}_{-0.01}$ |
| NGC 2634 | 4.54 | 2930.00 | 2.71 | $7.95^{+0.00}_{-0.19}$ | 11.00 | 0.36 | $-5.37^{+0.36}_{-0.00}$ |
| NGC 2960 | 2.59 | 810.00 | 2.25 | $7.06^{+0.17}_{-0.16}$ | 10.44 | 0.93 | $-4.73^{+0.00}_{-0.01}$ |
| NGC 3031 | 2.81 | 610.00 | 2.51 | $7.83^{+0.07}_{-0.11}$ | 10.16 | 0.64 | $-5.52^{+0.16}_{-0.11}$ |
| NGC 3079 | 0.52 | 470.00 | 0.93 | $6.38^{+0.13}_{-0.11}$ | 9.92 | 1.00 | $-7.89^{+0.14}_{-0.12}$ |
| NGC 3227 | 2.60 | 1830.00 | 0.79 | $7.88^{+0.14}_{-0.13}$ | 10.04 | 0.83 | $-6.75^{+0.18}_{-0.23}$ |
| NGC 3368 | 1.19 | 310.00 | 1.80 | $6.89^{+0.10}_{-0.08}$ | 9.81 | 0.99 | $-5.68^{+0.04}_{-0.04}$ |
| NGC 3393 | 1.14 | 430.00 | 1.76 | $7.49^{+0.16}_{-0.05}$ | 10.23 | 0.97 | $-5.77^{+0.01}_{-0.04}$ |
| NGC 3627 | 3.17 | 570.00 | 2.47 | $6.95^{+0.05}_{-0.05}$ | 9.74 | 0.91 | $-4.77^{+0.00}_{-0.00}$ |
| NGC 4151 | 2.24 | 570.00 | 2.26 | $7.68^{+0.58}_{-0.15}$ | 10.27 | 0.82 | $-5.36^{+0.14}_{-0.77}$ |
| NGC 4258 | 3.21 | 1540.00 | 1.52 | $7.60^{+0.01}_{-0.01}$ | 10.05 | 0.82 | $-5.78^{+0.01}_{-0.01}$ |
| NGC 4303 | 1.02 | 140.00 | 2.33 | $6.58^{+0.26}_{-0.07}$ | 9.42 | 1.00 | $-5.60^{+0.15}_{-0.04}$ |
| NGC 4388 | 0.89 | 1870.00 | −0.49 | $6.90^{+0.11}_{-0.11}$ | 10.07 | 1.00 | $-7.71^{+0.07}_{-0.07}$ |
| NGC 4458 | 10.10 | 4090.00 | 6.20 | $6.76^{+0.21}_{-0.08}$ | 10.27 | 0.48 | $-2.88^{+0.04}_{-0.09}$ |
| NGC 4478 | 3.11 | 1130.00 | 2.35 | $7.50^{+0.19}_{-0.01}$ | 10.56 | 0.80 | $-4.91^{+0.01}_{-0.16}$ |
| NGC 4501 | 2.33 | 1150.00 | 1.26 | $7.13^{+0.08}_{-0.08}$ | 10.11 | 0.95 | $-5.58^{+0.00}_{-0.00}$ |
| NGC 4736 | 0.93 | 210.00 | 2.21 | $6.78^{+0.11}_{-0.09}$ | 9.89 | 1.00 | $-5.64^{+0.06}_{-0.05}$ |
| NGC 4826 | 0.73 | 370.00 | 1.00 | $6.07^{+0.16}_{-0.14}$ | 9.55 | 1.00 | $-7.41^{+0.14}_{-0.12}$ |
| NGC 4945 | 3.40 | 470.00 | 2.56 | $6.15^{+0.30}_{-0.30}$ | 9.39 | 0.95 | $-4.83^{+0.08}_{-0.09}$ |
| NGC 5017 | 5.11 | 1910.00 | 3.67 | $7.98^{+0.00}_{-0.20}$ | 10.93 | 0.19 | $-4.96^{+0.40}_{-0.00}$ |
| NGC 5495 | 2.60 | 1760.00 | 1.35 | $7.04^{+0.09}_{-0.08}$ | 10.54 | 0.95 | $-5.31^{+0.01}_{-0.01}$ |
| NGC 5765 | 1.46 | 720.00 | 1.13 | $7.72^{+0.05}_{-0.05}$ | 10.04 | 0.95 | $-6.34^{+0.04}_{-0.04}$ |
| NGC 5831 | 4.72 | 3360.00 | 2.76 | $7.89^{+0.13}_{-0.02}$ | 11.08 | 0.39 | $-5.15^{+0.04}_{-0.25}$ |
| NGC 6264 | 1.04 | 840.00 | 0.60 | $7.51^{+0.06}_{-0.06}$ | 10.01 | 0.99 | $-6.70^{+0.00}_{-0.01}$ |
| NGC 6323 | 2.09 | 830.00 | 1.24 | $7.02^{+0.14}_{-0.13}$ | 9.86 | 0.97 | $-5.70^{+0.02}_{-0.02}$ |
| NGC 7582 | 2.20 | 510.00 | 2.25 | $7.67^{+0.08}_{-0.09}$ | 10.15 | 0.83 | $-5.39^{+0.08}_{-0.08}$ |
| UGC 3789 | 2.37 | 380.00 | 2.80 | $7.06^{+0.05}_{-0.05}$ | 10.18 | 0.92 | $-4.48^{+0.00}_{-0.00}$ |
| UGC 6093 | 1.55 | 1360.00 | 0.68 | $7.41^{+0.03}_{-0.04}$ | 10.35 | 0.97 | $-6.26^{+0.00}_{-0.00}$ |

**Note.** Table of stellar profile parameters for SMBHs obtained from H. Pfister et al. (2020). The parameters are (from left to right) the Sérsic index $n$, effective radius $R_{\mathrm{eff}}$ in pc, central density $\rho_0$, BH mass $M_{\mathrm{BH}}$, stellar mass $M_\star$, pinhole fraction $f$, and TDE rate $\Gamma$. Some galaxies are fitted with multiple components (NSCs or bulge), which are indicated by different lower indices (1, 2, or b).

## Appendix B
## Scaling of the Diffusion Coefficient

In this appendix, we show in detail how the diffusion coefficient scales with the stellar mass distribution. We summarize the derivations in Section 8.3 in J. J. Binney (2011) and Appendix B in H. Pfister et al. (2022). The orbital-averaged diffusion coefficient can be calculated by the following equation:

$$\bar{u}(E) = \frac{2}{P(E)} \int_{r_p}^{r_a} \frac{dr}{v_r} \lim_{R\to 0} \frac{\langle(\Delta R)^2\rangle}{2R}, \quad (B1)$$

where $P(E)$ is the orbital period, $v_r$ is the radial velocity of the orbit, and $r_a$ and $r_p$ are the apocenter and pericenter radius, respectively. The last term is the local diffusion coefficient,

defined as

$$\lim_{R \to 0} \frac{\langle (\Delta R)^2 \rangle}{2R} = \frac{32\pi^2 r^2 G^2 m_{\text{scatt}}^2 \ln\Lambda}{3 J_c(E)^2} \\ \times (3I_{1/2}(E) - I_{3/2}(E) + 2I_0(E)), \quad \text{(B2)}$$

which can be represented in terms of the moments of the stellar DF $f(E)$ (Equation (6)):

$$I_0(E) = \int_0^E f(E') dE \quad \text{(B3)}$$

$$I_{n/2}(E) = [2(\Psi - E)]^{-n/2} \int_E^\Psi [2(\Psi - E')]^{n/2} f(E') dE. \quad \text{(B4)}$$

We note that the diffusion coefficient is independent of the mass of the disrupted star, but instead depends on the mass of the background stars $m_{\text{scatt}}$. Therefore, when considering a distribution of stellar mass with an IMF $\phi(m)$, we take $m_{\text{scatt}}^2 = \langle m^2 \rangle = \int m^2 \phi(m) dm$. However, $f(E)$ itself is dependent on the stellar mass. Hence, we also need to scale it with $1/\langle m \rangle$. This means that the DF moments should also scale with $1/\langle m \rangle$. Combining these factors, we have the averaged orbital diffusion coefficient for a stellar population with distribution of stellar masses as

$$\bar{u}(E) = \frac{\langle m^2 \rangle}{\langle m \rangle} \bar{u}_{\text{mono}}(E). \quad \text{(B5)}$$

## Appendix C
### Effects of $M_{\text{BH}}$ on the TDE Pinhole Fraction

In this appendix, we demonstrate how $M_{\text{BH}}$ affects the TDE pinhole fraction $f$ using the example of an IMBH host galaxy NGC 1042. NGC 1042 has $M_{\text{BH,best}} = 2 \times 10^4 \, M_\odot$ and $M_{\text{BH,max}} = 3 \times 10^6 \, M_\odot$.

We plot the differential TDE rate $d^2\Gamma/dEd\ln\beta$ and the cumulative TDE rate fraction $\Gamma(E)/\Gamma_{\text{total}}$ in Figure C1 of NGC 1042 for both $M_{\text{BH,best}}$ and $M_{\text{BH,max}}$. One can see from the top panels that for the lower BH mass $M_{\text{BH,best}}$ most TDEs happen in the pinhole regime (to the left of the dashed line corresponding to $q = 1$) as opposed to the diffusive regime (to the right of the dashed line). This is simply because a lower $M_{BH}$ allows a higher maximum $\beta$ at which the star can still be disrupted before plunging into the BH event horizon, as demonstrated by the bottom panels. This higher cutoff value of $\beta$ increases the contribution of the pinhole regime, since the pinhole regime allows $\beta$ from large values, whereas the diffusive regime has $\beta \sim 1$.

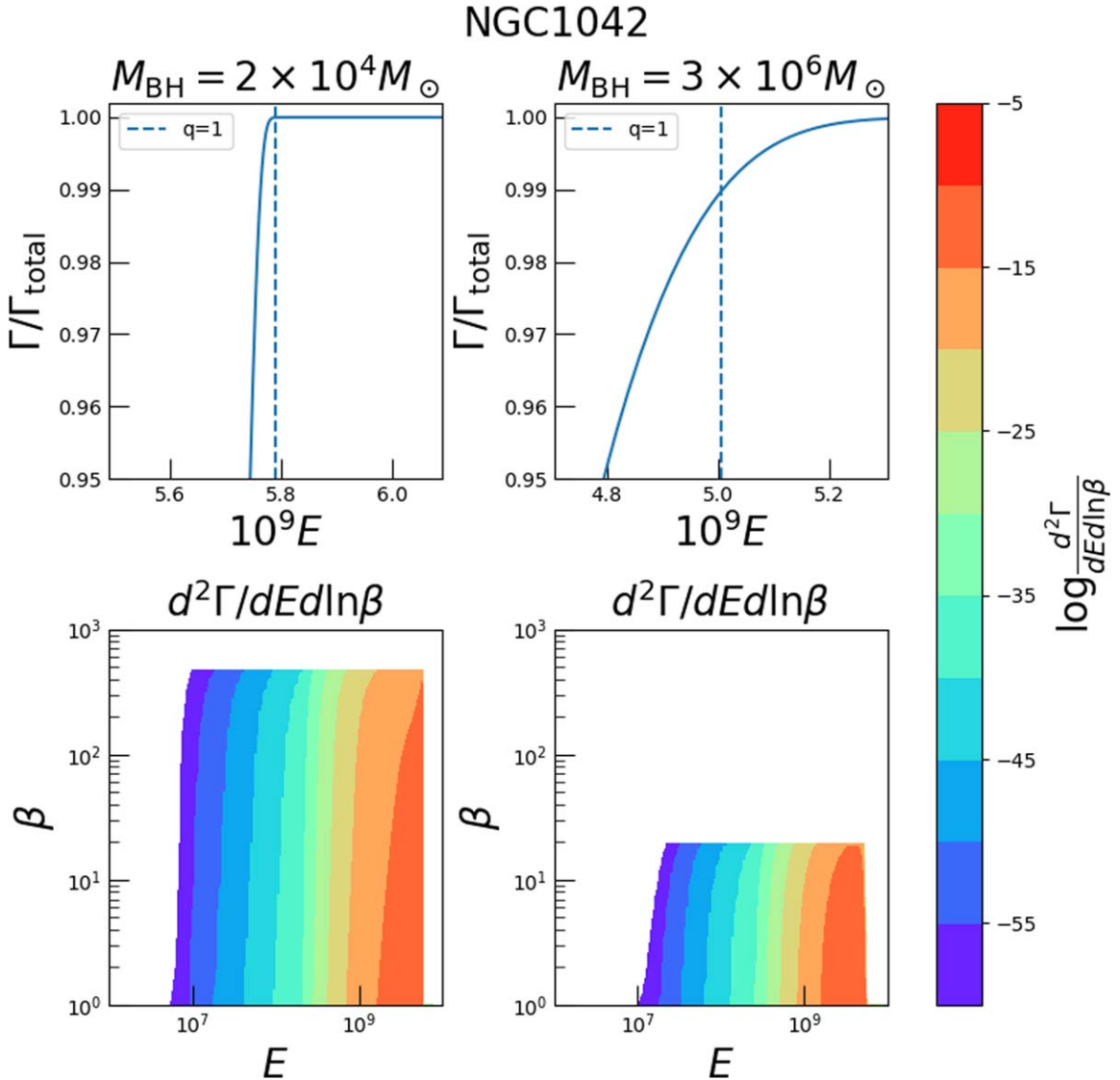

**Figure C1.** The TDE rate of an IMBH host galaxy NGC 1042 calculated using $M_{\text{BH,best}}$ (left) and $M_{\text{BH,max}}$ (right). The top row shows the cumulative TDE rate fraction as a function of $E$. The blue vertical dashed line indicates $q = 1$, which is the boundary separating the pinhole regime (left side) and diffusive regime (right side). The bottom row shows contour plots illustrating how the differential TDE rate $d\Gamma^2/dEd\ln\beta$ depends on $\beta$ (penetration parameter) and $E$ (orbital specific energy).

At the same time, one can also see that the boundary between the pinhole regime and the diffusive regime shifts to a higher $E$ for the lower $M_{BH}$. This is because $M_{BH}$ is related to the size of the loss cone $R_{lc}$ (Equation (11)). When $M_{BH}$ decreases, $R_{lc}$ also decreases, which in turn increases the $E$ at which the $q = 1$ boundary lies (Equation (12)). This also contributes to the higher pinhole fraction for the lower $M_{BH}$.

## Appendix D
## TDE Rates and Stellar Densities

In addition to $\Gamma$ versus $\rho_0$ shown in Figure 4, we show the correlation between $\Gamma$ and two more stellar densities in Figure D1: the stellar density at the BH influence radius $\rho_{inf}$, and the stellar density at 1pc $\rho_{1pc}$. One can see that the TDE rate generally increases with both $\rho_{inf}$ and $\rho_0$ as expected, for either the SMBH or IMBH host galaxies/clusters. However, there still exists some offset akin to the one shown in the $\Gamma$ versus $\rho_0$ correlation between the two populations. One possible contributor to this offset could be the disparity in the $R_{eff}$ between the two populations. Galaxies hosting SMBHs tend to have much larger $R_{eff}$ compare to those hosting IMBHs, resulting in lower central densities (which is evaluated at $R_{eff}$) for the former. We demonstrate that by evaluating the density at a specific radius (e.g., 1 pc), the offset is considerably reduced. Nevertheless, a slight offset between the two populations still persists.

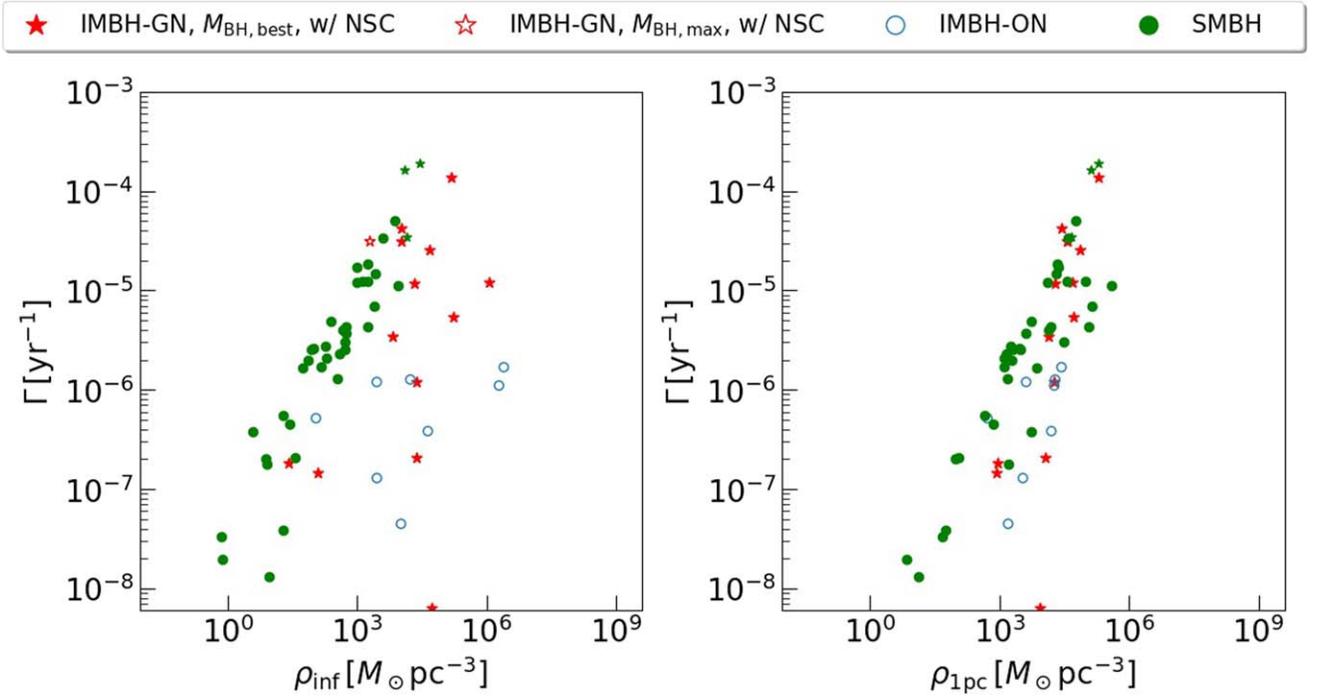

**Figure D1.** TDE rate $\Gamma$ vs. stellar density at the BH influence radius $\rho_{inf}$ (left), and the stellar density at 1 pc away from the galactic center $\rho_{1pc}$ (right). The symbol styles and colors are consistent with Figure 1(d).

## ORCID iDs

Janet N. Y. Chang ● https://orcid.org/0009-0004-2575-1924
Lixin Dai ● https://orcid.org/0000-0002-9589-5235
Hugo Pfister ● https://orcid.org/0000-0003-0841-5182
Rudrani Kar Chowdhury ● https://orcid.org/0000-0003-2694-933X
Priyamvada Natarajan ● https://orcid.org/0000-0002-5554-8896